\documentclass{aa}  
\usepackage{graphicx}
\usepackage{txfonts}
\usepackage{hyperref}

\begin{document}
   \title{New and updated stellar parameters for 71 evolved planet hosts\thanks{The data presented here are based on observations collected at the La Silla Paranal Observatory, ESO (Chile) with the FEROS spectrograph at the 2.2\,m telescope (ESO runs ID 70.C-0084, 088.C-0892, 089.C-0444, and 090.C-0146) and the HARPS spectrograph at the 3.6\,m telescope (ESO run ID 72.C-0488); at the Paranal Observatory, ESO (Chile) with the UVES spectrograph at the VLT Kueyen telescope (ESO runs ID 074.C-0134, 079.C-0131, 380.C-0083, and 083.C-0174); at the Spanish Observatorio del Roque de los Muchachos of the Instituto de Astrofisica de Canarias with the FIES spectrograph at the Nordic Optical Telescope, operated on the island of La Palma jointly by Denmark, Finland, Iceland, Norway, and Sweden (program ID 44-210); and at the Observatoire de Haute-Provence (OHP, CNRS/OAMP), France with the SOPHIE spectrographs at the 1.93\,m telescope (program ID 11B.DISC.SOUS).}}

   \subtitle{On the metallicity - giant planet connection}

   \author{A. Mortier\inst{1,2},
           N.C. Santos\inst{1,2},
           S.G. Sousa\inst{1,3},
           V.Zh. Adibekyan\inst{1},
           E. Delgado Mena\inst{1},
           M. Tsantaki\inst{1,2},
           G. Israelian\inst{3,4}
           \and
           M. Mayor\inst{5}
          }

   \institute{Centro de Astrof\'{\i}sica, Universidade do Porto, Rua das Estrelas, 4150-762 Porto, Portugal\\
              \email{amortier@astro.up.pt}
	      \and
	      Departamento de F\'{\i}sica e Astronomia, Faculdade de Ci\^encias, Universidade do Porto, Portugal	      
	\and
	      Instituto de Astrof\'{\i}sica de Canarias, 38200 La Laguna, Tenerife, Spain 
	\and
	      Departemento de Astrof\'{\i}sica, Universidad de La Laguna, E-38206 La Laguna, Tenerife, Spain
	\and
	      Observatoire de Gen\`eve, Universit\'e de Gen\`eve, 51 Ch. des Maillettes, 1290 Sauverny, Switzerland
	      }

   \date{Received 4 April 2013; Accepted 12 July 2013}

 
  \abstract
   {It is still being debated whether the well-known metallicity - giant planet correlation for dwarf stars is also valid for giant stars. For this reason, having precise metallicities is very important. Precise stellar parameters are also crucial to planetary research for several other reasons. Different methods can provide different results that lead to discrepancies in the analysis of planet hosts. }
   {To study the impact of different analyses on the metallicity scale for evolved stars, we compare different iron line lists to use in the atmospheric parameter derivation of evolved stars. Therefore, we use a sample of 71 evolved stars with planets. With these new homogeneous parameters, we revisit the metallicity - giant planet connection for evolved stars.}
   {A spectroscopic analysis based on Kurucz models in local thermodynamic equilibrium (LTE) was performed through the MOOG code to derive the atmospheric parameters. Two different iron line list sets were used, one built for cool FGK stars in general, and the other for giant FGK stars. Masses were calculated through isochrone fitting, using the Padova models. Kolmogorov-Smirnov tests (K-S tests) were then performed on the metallicity distributions of various different samples of evolved stars and red giants.}
   {All parameters compare well using a line list set, designed specifically for cool and solar-like stars to provide more accurate temperatures. All parameters derived with this line list set are preferred and are thus adopted for future analysis. We find that evolved planet hosts are more metal-poor than dwarf stars with giant planets. However, a bias in giant stellar samples that are searched for planets is present. Because of a colour cut-off, metal-rich low-gravity stars are left out of the samples, making it hard to compare dwarf stars with giant stars. Furthermore, no metallicity enhancement is found for red giants with planets ($\log g < 3.0$\,dex) with respect to red giants without planets.}
   {}


   \authorrunning{Mortier, A. et al.}
   \maketitle
%

\section{Introduction}

Since the discovery of the first extrasolar planet around a solar-like star in 1995 \citep[51 Peg b, ][]{Mayor95}, the search for extrasolar planetary systems has accelerated. Today, more than 900 planets have been announced \footnote{see \url{www.exoplanet.eu} for an updated table}. Most of them were detected using the very successful radial velocity technique or the photometric transit technique. This high number of known planetary systems enables us to perform a significant statistical analysis which can bring clarity to the theory of planet formation and evolution. Currently, there are still two main theories being debated in the literature, core-accretion \citep{Pol96,Mor09} and gravitational instability \citep{May02,Boss11}. 

Observational and theoretical evidence shows that the presence of a planet seems to depend on several stellar properties, such as mass and metallicity \citep{Udry07}. It has been well-established that dwarf stars with igher metallicity have a higher probability of harbouring a giant planet than their lower metallicity counterparts \citep{Gon97,San01,San04,Fis05,Sou11b,Me12}. This result is expected from the core-accretion models of planetary formation. Curiously, no such trend is observed for the lower mass planets. The Neptune-mass planets found so far seem to have a rather flat metallicity distribution \citep{Udry07,Sou08,Sou11b,Mayor11,Buch12}. It has also been suggested that intermediate-mass giant stars with planets have a different trend in metallicity \citep{Pas07,Ghe10b,Hek07}.

There is also growing evidence that stellar mass may play a role in the formation of giant planets. Giant planet frequency around low-mass M dwarfs is much lower than the one found for FGK dwarfs \citep{Bon05,Endl06,Nev13}. For higher mass stars, on the other hand, the frequency of giant planets seems to be higher \citep{Lov07,John07}.

To address these issues in a consistent way, it is important that a correct determination of stellar parameters is performed. In addition, a uniform derivation of the stellar parameters is a must for a thorough statistical analysis. 

The determination of stellar parameters for evolved stars, and especially the cool giant stars, has been debated in the literature for several years \citep[e.g. ][]{Hek07,Coh08,San09,San10,Ghe10b}. Several authors raise doubts about the metallicity scale in these evolved stars. This has lead to significant discrepancies in the different studies of giant and sub-giant samples. In this work, we compare two iron line lists to derive stellar parameters for giant stars with planets. We show that using different line lists will lead to different results. The impact on the metallicity - giant planet connection is then presented and discussed. 

In Sect. \ref{Dat}, we present the sample that has been used. Section \ref{Ana} describes the spectroscopic analysis that was used, together with the different line lists. Results are given in Sect. \ref{Res}. The effect on planet frequency as a function of metallicity is discussed in Sect. \ref{Freq}. We give a discusion in Sect. \ref{Disc} and we conclude in Sect. \ref{Con}.


\section{The sample}\label{Dat}

For this analysis, we used a sample of 71 stars. All these stars are of spectral type F, G, or K and are known to be orbited by a planet (according to the online catalogue \url{www.exoplanet.eu}). They were selected on the basis of their surface gravity, derived with the line list of \citet{Sou08}, see next section: $\log g<4.0$. Twenty of these stars were previously analysed with the same method presented in this work and their parameters were published by members of our team. The references can be found in Table \ref{TabLog}. For the 51 remaining stars, we gathered optical spectra through observations made by our team, and with the use of the ESO archive.

\begin{longtab}
\begin{longtable}{lcc}
\caption{\label{TabLog} Data log}\\
\hline\hline
Name & Instrument & Reference \\
\hline
\endfirsthead
\caption{continued.}\\
\hline\hline
Name & Instrument & Reference \\
\hline
\endhead
\hline
\endfoot
\object{11Com} & UVES & This work \\
\object{18Del} & UVES & This work \\
\object{24Sex} & FEROS & This work \\
\object{7CMa} & FEROS & This work \\
\object{75Cet} & FEROS & This work \\
\object{81Cet} & FEROS & This work \\
\object{91Aqr} & UVES & \citet{San05} \\
\object{alfAri} & FEROS & This work \\
\object{BD+202457} & FEROS & This work \\
\object{BD+20274} & FEROS & This work \\
\object{BD+48738} & SOPHIE & This work \\
\object{epsCrB} & UVES & This work \\
\object{epsTau} & UVES & This work \\
\object{gam01Leo} & FEROS & This work \\
\object{gammaCephei} & SARG & \citet{San04} \\
\object{HD100655} & FEROS & This work \\
\object{HD102272} & UVES & This work \\
\object{HD102329} & FEROS & This work \\
\object{HD104985} & SARG & \citet{San05} \\
\object{HD106270} & FIES & This work \\
\object{HD108863} & FEROS & This work \\
\object{HD110014} & FEROS & This work \\
\object{HD116029} & FEROS & This work \\
\object{HD11964} & FEROS & \citet{Sou06} \\
\object{HD11977} & CORALIE & \citet{Sou06} \\
\object{HD122430} & FEROS & This work \\
\object{HD13189} & SARG & \citet{Sou06} \\
\object{HD148427} & FEROS & This work \\
\object{HD1502} & FEROS & This work \\
\object{HD156411} & HARPS & \citet{Sou11b} \\
\object{HD159868} & HARPS & \citet{Sou08} \\
\object{HD167042} & SOPHIE & This work \\
\object{HD1690} & HARPS & \citet{Mou11a} \\
\object{HD171028} & HARPS & \citet{Sou11b} \\
\object{HD175541} & UVES & This work \\
\object{HD177830} & SARG & \citet{San04} \\
\object{HD180902} & FEROS & This work \\
\object{HD181342} & FEROS & This work \\
\object{HD18742} & FEROS & This work \\
\object{HD192699} & UVES & This work \\
\object{HD200964} & FEROS & This work \\
\object{HD206610} & FEROS & This work \\
\object{HD210702} & UVES & This work \\
\object{HD212771} & FEROS & This work \\
\object{HD27442} & FEROS & \citet{San04} \\
\object{HD28678} & FEROS & This work \\
\object{HD30856} & FEROS & This work \\
\object{HD33142} & FEROS & This work \\
\object{HD38529} & FEROS & \citet{San04} \\
\object{HD38801} & FEROS & This work \\
\object{HD4313} & FEROS & This work \\
\object{HD47536} & FEROS & \citet{San04} \\
\object{HD48265} & UVES & This work \\
\object{HD5319} & UVES & This work \\
\object{HD5608} & FEROS & This work \\
\object{HD5891} & FEROS & This work \\
\object{HD59686} & UVES & \citet{San05} \\
\object{HD62509} & UVES & This work \\
\object{HD66141} & UVES & This work \\
\object{HD73534} & FEROS & This work \\
\object{HD88133} & UVES & \citet{San05} \\
\object{HD89744} & UES & \citet{San04} \\
\object{HD95089} & FEROS & This work \\
\object{HD96063} & FEROS & This work \\
\object{HD98219} & FEROS & This work \\
\object{HIP75458} & SARG & \citet{San04} \\
\object{kappaCrB} & UVES & This work \\
\object{ksiAql} & UVES & This work \\
\object{NGC2423 No3} & UVES & \citet{San09} \\
\object{NGC4349 No127} & UVES & \citet{San09} \\
\object{nuOph} & FEROS & This work \\

\end{longtable}
\end{longtab}

\begin{table}
\caption{Spectrograph details: resolving power and spectral ranges.}
\label{TabINS}
\centering
\begin{tabular}{cccc}
\hline\hline
Instrument & Resolving power & Spectral range & Stars \\
 & $\lambda/\Delta\lambda$ & \AA & \\
\hline
UVES & 110000 & 3000 - 6800 & 19 \\
FEROS & 48000 & 3600 - 9200 & 38 \\
HARPS & 100000 & 3800 - 7000 & 4 \\
CORALIE & 50000 & 3800 - 6800 & 1 \\
SOPHIE & 75000 & 3820 - 6920 & 2 \\
SARG & 57000 - 86000 & 5100 - 10100 & 5 \\
FIES & 67000 & 3700 - 7300 & 1 \\
UES & 55000 & 4000 - 10000 & 1 \\
\hline
\end{tabular}
\end{table}

In total, eight different spectrographs were used (see Table \ref{TabINS}): UVES (VLT Kueyen telescope, Paranal, Chile); FEROS (2.2m ESO/MPI telescope, La Silla, Chile); HARPS (3.6m ESO telescope, La Silla, Chile); CORALIE (1.2m Swiss telescope, La Silla, Chile); SOPHIE (1.93m telescope, OHP, France); SARG (TNG Telescope, La Palma, Spain); FIES (Nordic Optical Telescope, La Palma, Spain); and UES (William Herschel Telescope, La Palma, Spain). The spectra were reduced using the available pipelines and IRAF \footnote{IRAF is distributed by National Optical Astronomy Observatories, operated by the Association of Universities for Research in Astronomy, Inc., under contract with the National Science Foundation, USA.}. The spectra were corrected for radial velocity with the IRAF task \texttt{DOPCOR}. Individual exposures of multiple observed stars with the same instrument, were added using the task \texttt{SCOMBINE} in IRAF. The data log can be found in Table \ref{TabLog}.

The usage of different spectrographs and different pipelines can introduce systematic offsets in the analysis of the data. However, \citet{San04} show that the spectroscopic analysis used in this work is not significantly affected by the use of different spectrographs. They observed several stars with different spectrographs and the resulting atmospheric parameters were all similar and within the 1-$\sigma$ errorbars.

\section{Spectroscopic analysis}\label{Ana}

\subsection{Technique}

From the spectra, we derived atmospheric stellar parameters: the effective temperature T$_{eff}$, the surface gravity $\log g$, the metallicity [Fe/H] and the microturbulence $\xi$. The procedure we followed is described in \citet{San04}, and is based on the equivalent widths of \ion{Fe}{i} and \ion{Fe}{ii} lines, and iron excitation and ionization equilibrium, assumed in local thermodynamic equilibrium (LTE). Therefore, the 2002 version of MOOG\footnote{\url{http://www.as.utexas.edu/~chris/moog.html}} \citep{Sne73} and a grid of ATLAS plane-parallel model atmospheres \citep{Kur93} are used. This, mostly automatic, analysis was also used for the 20 stars whose parameters were previously published (see Table \ref{TabLog}).

To measure the equivalent widths (EWs) of the iron lines, the code ARES is used \citep[Automatic Routine for line Equivalent widths in stellar Spectra -][]{Sou07}. The input parameters for ARES, are the same as in \citet[hereafter SO08]{Sou08}, except for the \emph{ rejt} parameter, which determines the calibration of the continuum position. Since this parameter strongly depends on the signal-to-noise (S/N) of the spectra, different values are needed. A uniform S/N value is derived for the spectra with the IRAF routine \texttt{BPLOT}. Therefore, three spectral regions are used: [5744\AA, 5747\AA], [6047\AA, 6053\AA], and [6068\AA, 6076\AA]. 

Then, the \emph{ rejt} parameter was set by eye for ten spectra with different S/N (representable for the whole sample). Afterwards, all the \emph{ rejt} parameters were derived by a simple interpolation of these values. This method ensures a uniform usage of the \emph{ rejt} parameter, since we otherwise do not have access to a uniform source for the S/N as they do in SO08. Table \ref{TabRejt} lists the dependence of the \emph{ rejt} parameter to the S/N. 

\begin{table}
\caption{Dependence of the \emph{ rejt} parameter on the S/N.}
\label{TabRejt}
\centering
\begin{tabular}{rc|rc}
\hline\hline
S/N condition & \emph{ rejt} & S/N condition & \emph{ rejt} \\
\hline
S/N $\leq$ 20 & 0.989 & 80 $<$ S/N $\leq$ 120 & 0.994 \\
20 $<$ S/N $\leq$ 30 & 0.990 & 120 $<$ S/N $\leq$ 160 & 0.995 \\
30 $<$ S/N $\leq$ 40 & 0.991 & 160 $<$ S/N $\leq$ 200 & 0.996 \\
40 $<$ S/N $\leq$ 60 & 0.992 & S/N $>$ 200 & 0.997 \\
60 $<$ S/N $\leq$ 80 & 0.993 & & \\
\hline
\end{tabular}
\end{table}

We performed the spectroscopic analysis in LTE. Stellar parameters, especially metallicity, can differ if non-LTE effects are considered. Results obtained with classical models depend on the choice of the line list and, in particular, on the ionization balance of \ion{Fe}{i}/\ion{Fe}{ii} \citep{Ber12}. In a recent study, \citet{Lin12} quantified the corrections between an LTE and a non-LTE analysis. They show that the largest impact is seen for very metal-poor and hot giant stars. Since our sample consists of cool giant stars with metallicities higher than $-1.0$\,dex, our assumption of LTE is justified.

\subsection{Different line lists}

The iron lines that are used to derive atmospheric parameters should be chosen carefully so that the equivalent widths can be measured accurately. Ideal lines are single and not blended or saturated. Different line lists exist in the literature, some very general, some for specific types of stars. To find a reliable line list for evolved stars, in this work we used two different Fe line list sets to measure the equivalent widths and then derived the atmospheric parameters. 

First, all stars were analysed with the large line list taken from SO08, as our team usually does. For some stars that were analysed before 2008, the smaller line list from \citet{San04} was used. In SO08, the authors show that the stellar parameters, obtained with both lists, compare very well. Since more lines allow for smaller errorbars, the large SO08 line list is preferred. For cool stars (below 5200K), however, the results from using these line lists were unsatisfactory. The derived temperatures were too high, when compared with other methods like the InfraRed flux Method \citep{Sou08}. Therefore, a new line list was built, specifically for these cooler stars \citep[][hereafter TS13]{Tsa13}, based on the SO08 line list. Only weak and isolated lines were left, since blending effects play a huge role in cool stars. The authors show that their new temperatures are in very good agreement with the results from the InfraRed flux Method. The other atmospheric parameters are comparable with the results from using the SO08 line list. Since most giant stars in our sample (61 out of 71) have temperatures lower than 5200K, we used this line list for these 61 stars. For the remaining 10 stars, we still used the SO08 line list. Hereafter we will refer to this line list set as TS13 - SO08.

Second, we also derived the parameters by making use of the small line list (20 \ion{Fe}{i} and 6 \ion{Fe}{ii} lines) from \citet[][hereafter HM07]{Hek07}. This line list was made specifically for giant stars where all lines were carefully selected to avoid blends by atomic and CN lines. The reference solar iron abundance used to make this list is $A(Fe)_{\odot} = 7.49$. For the line lists of SO08 and TS13, the reference value of $A(Fe)_{\odot} = 7.47$ was used.

\subsection{Reference star Arcturus}

\begin{table*}
\caption{Atmospheric parameters for the reference giant star Arcturus.}
\label{TabArc}
\centering
\begin{tabular}{ccccc}
\hline\hline
 & T$_{eff}$ & $\log g_{spec}$ & [Fe/H] & $\xi$ \\
 & (K) & (cm s$^{-2}$) & (dex) & (km s$^{-1}$) \\
\hline
This work - SO08 & 4392 $\pm$ 56 & 1.92 $\pm$ 0.13 & -0.57 $\pm$ 0.03 & 1.90 $\pm$ 0.05 \\
This work - TS13 & 4368 $\pm$ 63 & 1.86 $\pm$ 0.19 & -0.59 $\pm$ 0.04 & 1.86 $\pm$ 0.07 \\
This work - HM07 & 4411 $\pm$ 104 & 2.30 $\pm$ 0.21 & -0.56 $\pm$ 0.07 & 2.08 $\pm$ 0.15 \\
MARCS models - TS13 & 4408 $\pm$ 70 & 1.89 $\pm$ 0.19 & -0.55 $\pm$ 0.04 & 1.89 $\pm$ 0.06 \\
\citet{Ram11} & 4286 $\pm$ 30 & 1.66 $\pm$ 0.05 & -0.52 $\pm$ 0.04 & 1.74 \\
PASTEL mean & 4317 $\pm$ 63 & 1.68 $\pm$ 0.32 & -0.54 $\pm$ 0.11 & ~ \\
\hline
\end{tabular}
\end{table*}

To test our line lists, we derived parameters for the K giant Arcturus. This star is an excellent reference star for giant studies since it is very bright and easy to observe with different telescopes. Furthermore, it has been studied multiple times, so there is a lot of data for comparison. For our analysis, we used a spectrum from the archive from the NARVAL spectrograph (with a resolution of 70000) at the 2m Bernard Lyot Telescope in Toulouse, France.

Table \ref{TabArc} lists our results derived using the three different line lists mentioned above. These all compare remarkebly well. In Fig. \ref{FigComp}, this reference star is overplotted. We also list the values from a recent study by \citet{Ram11}. Furthermore, we take the mean and standard deviation of all values listed in the PASTEL catalogue \citep{Soub10}. All values compare within the errorbars. We are thus confident that our line lists are suited for analysing cool giant stars.

The agreement in temperature between the line lists of SO08 and TS13 is remarkable, since the SO08 line list normally results in overestimated temperatures. This star is so bright that the spectrum is of very good quality, with both high resolution and high signal to noise. The SO08 line list is thus probably less affected by blended lines resulting in compatible results with the TS13 line list.

\subsection{Stellar masses}

Stellar masses were estimated as in previous works \citep[e.g.][]{San04,Sou11b}. Stellar evolutionary models from the Padova group were used, through their webpage\footnote{\url{http://stev.oapd.inaf.it/cgi-bin/param}} \citep{Das06}. This web interface needs four parameters. For the temperature and metallicity, the values from our analysis were used. For the V magnitude and parallax, the Hipparcos values were used \citep{VanL07} whenever possible. If no errors were provided in the database, the typical values $0.05$ and $1.0$\,mas, respectively, were used.

If no V magnitude was present in the Hipparcos database, we used the value presented by Simbad. If the Hipparcos database did not provide a parallax or if the parallax error was larger than $10\%$ of the value, an iterative method was used to get the masses. This method was previously introduced in \citet{San10} and makes use of the bolometric correction from \citet{Flo96} and the Padova web interface.

\section{Stellar parameters}\label{Res}

Table \ref{TabParMar} lists all stellar parameters derived with the TS13 line list for stars cooler than 5200\,K, and SO08 for stars with a temperature above this value (the TS13 - SO08 line list set). For the cooler stars, we list the SO08 results in Appendix \ref{ApSou} (Table \ref{TabParSou}). In Table \ref{TabParMel}, we list the parameters derived with the HM07 line list.

\begin{longtab}
\begin{longtable}{lccccc}
\caption{\label{TabParMar} Stellar parameters derived with the TS13 line list for the stars cooler than 5200\,K, and the SO08 line list for the hotter stars (marked with $\ast$). We adopt these parameters for all future analyses of these stars.}\\
\hline\hline
Name & T$_{eff}$ & $\log g_{spec}$ & [Fe/H] & $\xi$ & M$_{\ast}$ \\
 & (K) & (cm s$^{-2}$) & (dex) & (km s$^{-1}$)  & (M$_{\odot}$)  \\
\hline
\endfirsthead
\caption{continued.}\\
\hline\hline
Name & T$_{eff}$ & $\log g_{spec}$ & [Fe/H] & $\xi$ & M$_{\ast}$ \\
 & (K) & (cm s$^{-2}$) & (dex) & (km s$^{-1}$)  & (M$_{\odot}$)  \\
\hline
\endhead
\hline
\endfoot
11Com & 4830 $\pm$ 79 & 2.61 $\pm$ 0.13 & -0.34 $\pm$ 0.06 & 1.70 $\pm$ 0.10 & 2.04 $\pm$ 0.29 \\
18Del & 5076 $\pm$ 38 & 3.08 $\pm$ 0.10 & 0.00 $\pm$ 0.03 & 1.32 $\pm$ 0.04 & 2.33 $\pm$ 0.05 \\
24Sex & 5069 $\pm$ 62 & 3.40 $\pm$ 0.13 & -0.01 $\pm$ 0.05 & 1.27 $\pm$ 0.07 & 1.81 $\pm$ 0.08 \\
75Cet & 4904 $\pm$ 47 & 2.87 $\pm$ 0.14 & 0.02 $\pm$ 0.04 & 1.41 $\pm$ 0.05 & 2.15 $\pm$ 0.18 \\
7CMa & 4761 $\pm$ 79 & 3.11 $\pm$ 0.22 & 0.21 $\pm$ 0.05 & 1.19 $\pm$ 0.08 & 1.34 $\pm$ 0.13 \\
81Cet & 4845 $\pm$ 51 & 2.45 $\pm$ 0.19 & -0.07 $\pm$ 0.04 & 1.58 $\pm$ 0.05 & 1.74 $\pm$ 0.21 \\
91Aqr & 4681 $\pm$ 92 & 2.65 $\pm$ 0.22 & 0.00 $\pm$ 0.05 & 1.59 $\pm$ 0.09 & 1.32 $\pm$ 0.23 \\
alfAri & 4513 $\pm$ 72 & 2.49 $\pm$ 0.21 & -0.16 $\pm$ 0.03 & 1.50 $\pm$ 0.07 & 1.33 $\pm$ 0.22 \\
BD+202457 & 4259 $\pm$ 64 & 1.77 $\pm$ 0.19 & -0.79 $\pm$ 0.03 & 1.98 $\pm$ 0.07 & 1.06 $\pm$ 0.21 \\
BD+20274 & 4391 $\pm$ 79 & 2.03 $\pm$ 0.21 & -0.32 $\pm$ 0.04 & 1.80 $\pm$ 0.08 & 1.17 $\pm$ 0.23 \\
BD+48738 & 4658 $\pm$ 118 & 2.62 $\pm$ 0.26 & -0.09 $\pm$ 0.08 & 1.70 $\pm$ 0.12 & 1.19 $\pm$ 0.24 \\
epsCrB & 4436 $\pm$ 56 & 1.94 $\pm$ 0.15 & -0.22 $\pm$ 0.03 & 1.68 $\pm$ 0.06 & 1.44 $\pm$ 0.18 \\
epsTau & 4946 $\pm$ 70 & 2.62 $\pm$ 0.15 & 0.17 $\pm$ 0.06 & 1.48 $\pm$ 0.07 & 2.73 $\pm$ 0.10 \\
gam01Leo & 4428 $\pm$ 53 & 1.97 $\pm$ 0.17 & -0.41 $\pm$ 0.03 & 1.74 $\pm$ 0.05 & 1.57 $\pm$ 0.18 \\
gammaCephei & 4764 $\pm$ 112 & 3.10 $\pm$ 0.27 & 0.13 $\pm$ 0.06 & 1.06 $\pm$ 0.12 & 1.26 $\pm$ 0.14 \\
HD100655 & 4801 $\pm$ 60 & 2.81 $\pm$ 0.18 & -0.02 $\pm$ 0.04 & 1.50 $\pm$ 0.06 & 1.71 $\pm$ 0.33 \\
HD102329 & 4745 $\pm$ 71 & 2.96 $\pm$ 0.16 & 0.05 $\pm$ 0.04 & 1.39 $\pm$ 0.07 & 1.30 $\pm$ 0.15 \\
HD102272 & 4807 $\pm$ 34 & 2.57 $\pm$ 0.13 & -0.38 $\pm$ 0.03 & 1.53 $\pm$ 0.04 & 1.10 $\pm$ 0.20 \\
HD104985 & 4819 $\pm$ 161 & 2.96 $\pm$ 0.27 & -0.35 $\pm$ 0.11 & 2.00 $\pm$ 0.22 & 1.04 $\pm$ 0.27 \\
HD106270$^{\ast}$ & 5601 $\pm$ 24 & 3.72 $\pm$ 0.05 & 0.06 $\pm$ 0.02 & 1.32 $\pm$ 0.02 & 1.33 $\pm$ 0.05 \\
HD108863 & 4919 $\pm$ 44 & 2.99 $\pm$ 0.13 & 0.02 $\pm$ 0.03 & 1.31 $\pm$ 0.04 & 2.08 $\pm$ 0.14 \\
HD110014 & 4478 $\pm$ 106 & 2.52 $\pm$ 0.31 & 0.14 $\pm$ 0.05 & 1.90 $\pm$ 0.11 & 2.09 $\pm$ 0.39 \\
HD116029 & 4811 $\pm$ 57 & 3.21 $\pm$ 0.17 & 0.08 $\pm$ 0.04 & 1.11 $\pm$ 0.06 & 1.33 $\pm$ 0.11 \\
HD11964$^{\ast}$ & 5372 $\pm$ 35 & 3.99 $\pm$ 0.04 & 0.14 $\pm$ 0.05 & 1.09 $\pm$ 0.04 & 1.08 $\pm$ 0.02 \\
HD11977 & 5067 $\pm$ 42 & 2.91 $\pm$ 0.15 & -0.16 $\pm$ 0.04 & 1.59 $\pm$ 0.05 & 2.31 $\pm$ 0.12 \\
HD122430 & 4474 $\pm$ 102 & 2.43 $\pm$ 0.29 & -0.09 $\pm$ 0.05 & 1.81 $\pm$ 0.1 & 1.68 $\pm$ 0.32 \\
HD13189$^{\ast\ast}$ & 4337 $\pm$ 133 & 1.83 $\pm$ 0.31 & -0.39 $\pm$ 0.19 & 1.99 $\pm$ 0.12 & 1.17 $\pm$ 0.23 \\
HD148427 & 4962 $\pm$ 45 & 3.39 $\pm$ 0.12 & 0.03 $\pm$ 0.03 & 1.06 $\pm$ 0.05 & 1.36 $\pm$ 0.06 \\
HD1502 & 4984 $\pm$ 46 & 3.30 $\pm$ 0.10 & -0.04 $\pm$ 0.03 & 1.21 $\pm$ 0.05 & 1.47 $\pm$ 0.11 \\
HD156411$^{\ast}$ & 5910 $\pm$ 16 & 3.99 $\pm$ 0.01 & -0.11 $\pm$ 0.01 & 1.31 $\pm$ 0.01 & 1.24 $\pm$ 0.02 \\
HD159868$^{\ast}$ & 5558 $\pm$ 15 & 3.96 $\pm$ 0.02 & -0.08 $\pm$ 0.01 & 1.02 $\pm$ 0.01 & 1.16 $\pm$ 0.04 \\
HD167042 & 5028 $\pm$ 53 & 3.35 $\pm$ 0.18 & 0.03 $\pm$ 0.04 & 1.26 $\pm$ 0.06 & 1.63 $\pm$ 0.06 \\
HD1690 & 4364 $\pm$ 111 & 2.16 $\pm$ 0.27 & -0.29 $\pm$ 0.06 & 1.75 $\pm$ 0.12 & 1.18 $\pm$ 0.23 \\
HD171028$^{\ast}$ & 5671 $\pm$ 16 & 3.84 $\pm$ 0.03 & -0.48 $\pm$ 0.01 & 1.24 $\pm$ 0.02 & 1.01 $\pm$ 0.06 \\
HD175541 & 5111 $\pm$ 38 & 3.56 $\pm$ 0.08 & -0.11 $\pm$ 0.03 & 1.13 $\pm$ 0.05 & 1.34 $\pm$ 0.08 \\
HD177830 & 4752 $\pm$ 79 & 3.37 $\pm$ 0.20 & 0.30 $\pm$ 0.05 & 0.99 $\pm$ 0.12 & 1.17 $\pm$ 0.10 \\
HD180902 & 5040 $\pm$ 47 & 3.41 $\pm$ 0.18 & 0.01 $\pm$ 0.04 & 1.15 $\pm$ 0.05 & 1.53 $\pm$ 0.09 \\
HD181342 & 4965 $\pm$ 56 & 3.27 $\pm$ 0.19 & 0.15 $\pm$ 0.04 & 1.25 $\pm$ 0.06 & 1.70 $\pm$ 0.09 \\
HD18742 & 5016 $\pm$ 32 & 3.15 $\pm$ 0.08 & -0.15 $\pm$ 0.03 & 1.23 $\pm$ 0.04 & 1.73 $\pm$ 0.19 \\
HD192699 & 5141 $\pm$ 20 & 3.45 $\pm$ 0.07 & -0.20 $\pm$ 0.02 & 1.11 $\pm$ 0.02 & 1.58 $\pm$ 0.04 \\
HD200964 & 5082 $\pm$ 38 & 3.41 $\pm$ 0.08 & -0.20 $\pm$ 0.03 & 1.28 $\pm$ 0.05 & 1.57 $\pm$ 0.06 \\
HD206610 & 4821 $\pm$ 55 & 3.24 $\pm$ 0.16 & 0.10 $\pm$ 0.03 & 1.14 $\pm$ 0.06 & 1.30 $\pm$ 0.12 \\
HD210702 & 5000 $\pm$ 44 & 3.36 $\pm$ 0.08 & 0.04 $\pm$ 0.03 & 1.15 $\pm$ 0.05 & 1.71 $\pm$ 0.06 \\
HD212771 & 5091 $\pm$ 39 & 3.38 $\pm$ 0.07 & -0.14 $\pm$ 0.03 & 1.20 $\pm$ 0.04 & 1.51 $\pm$ 0.08 \\
HD27442 & 4781 $\pm$ 76 & 3.46 $\pm$ 0.19 & 0.33 $\pm$ 0.05 & 1.09 $\pm$ 0.11 & 1.24 $\pm$ 0.10 \\
HD28678 & 5052 $\pm$ 29 & 3.07 $\pm$ 0.09 & -0.21 $\pm$ 0.02 & 1.31 $\pm$ 0.03 & 2.03 $\pm$ 0.20 \\
HD30856 & 4973 $\pm$ 29 & 3.15 $\pm$ 0.11 & -0.14 $\pm$ 0.02 & 1.19 $\pm$ 0.03 & 1.36 $\pm$ 0.07 \\
HD33142 & 5049 $\pm$ 41 & 3.34 $\pm$ 0.14 & 0.03 $\pm$ 0.03 & 1.17 $\pm$ 0.04 & 1.62 $\pm$ 0.09 \\
HD38529$^{\ast}$ & 5674 $\pm$ 40 & 3.94 $\pm$ 0.12 & 0.40 $\pm$ 0.06 & 1.38 $\pm$ 0.05 & 1.34 $\pm$ 0.02 \\
HD38801$^{\ast}$ & 5314 $\pm$ 43 & 3.82 $\pm$ 0.08 & 0.25 $\pm$ 0.03 & 1.29 $\pm$ 0.06 & 1.22 $\pm$ 0.07 \\
HD4313 & 4966 $\pm$ 40 & 3.36 $\pm$ 0.17 & 0.05 $\pm$ 0.03 & 1.17 $\pm$ 0.05 & 1.53 $\pm$ 0.09 \\
HD47536 & 4447 $\pm$ 70 & 2.26 $\pm$ 0.17 & -0.65 $\pm$ 0.04 & 1.70 $\pm$ 0.07 & 0.98 $\pm$ 0.08 \\
HD48265$^{\ast}$ & 5798 $\pm$ 29 & 3.95 $\pm$ 0.14 & 0.36 $\pm$ 0.02 & 1.36 $\pm$ 0.03 & 1.21 $\pm$ 0.04 \\
HD5319 & 4869 $\pm$ 51 & 3.22 $\pm$ 0.10 & 0.02 $\pm$ 0.03 & 1.11 $\pm$ 0.05 & 1.28 $\pm$ 0.10 \\
HD5608 & 4911 $\pm$ 51 & 3.25 $\pm$ 0.16 & 0.12 $\pm$ 0.03 & 1.18 $\pm$ 0.06 & 1.66 $\pm$ 0.08 \\
HD5891 & 4825 $\pm$ 47 & 2.62 $\pm$ 0.10 & -0.38 $\pm$ 0.04 & 1.69 $\pm$ 0.05 & 1.09 $\pm$ 0.19 \\
HD59686 & 4666 $\pm$ 76 & 2.57 $\pm$ 0.21 & 0.12 $\pm$ 0.04 & 1.49 $\pm$ 0.07 & 2.02 $\pm$ 0.29 \\
HD62509 & 4935 $\pm$ 49 & 2.91 $\pm$ 0.13 & 0.09 $\pm$ 0.04 & 1.34 $\pm$ 0.05 & 2.08 $\pm$ 0.09 \\
HD66141 & 4320 $\pm$ 50 & 1.90 $\pm$ 0.15 & -0.42 $\pm$ 0.03 & 1.59 $\pm$ 0.05 & 1.00 $\pm$ 0.06 \\
HD73534 & 4884 $\pm$ 63 & 3.59 $\pm$ 0.20 & 0.16 $\pm$ 0.04 & 1.00 $\pm$ 0.08 & 1.17 $\pm$ 0.07 \\
HD88133$^{\ast}$ & 5438 $\pm$ 34 & 3.94 $\pm$ 0.11 & 0.33 $\pm$ 0.05 & 1.16 $\pm$ 0.03 & 1.18 $\pm$ 0.06 \\
HD89744$^{\ast}$ & 6234 $\pm$ 45 & 3.98 $\pm$ 0.05 & 0.22 $\pm$ 0.05 & 1.62 $\pm$ 0.08 & 1.48 $\pm$ 0.02 \\
HD95089 & 4950 $\pm$ 68 & 3.32 $\pm$ 0.18 & 0.00 $\pm$ 0.05 & 1.27 $\pm$ 0.08 & 1.37 $\pm$ 0.12 \\
HD96063 & 5131 $\pm$ 26 & 3.44 $\pm$ 0.06 & -0.20 $\pm$ 0.02 & 1.14 $\pm$ 0.03 & 1.41 $\pm$ 0.09 \\
HD98219 & 5046 $\pm$ 71 & 3.59 $\pm$ 0.14 & 0.05 $\pm$ 0.05 & 1.07 $\pm$ 0.08 & 1.62 $\pm$ 0.11 \\
HIP75458 & 4528 $\pm$ 111 & 2.49 $\pm$ 0.26 & -0.16 $\pm$ 0.06 & 1.64 $\pm$ 0.12 & 1.14 $\pm$ 0.16 \\
kappaCrB & 4876 $\pm$ 46 & 3.15 $\pm$ 0.14 & 0.13 $\pm$ 0.03 & 1.10 $\pm$ 0.06 & 1.58 $\pm$ 0.08 \\
ksiAql & 4714 $\pm$ 49 & 2.53 $\pm$ 0.11 & -0.27 $\pm$ 0.04 & 1.55 $\pm$ 0.05 & 1.11 $\pm$ 0.25 \\
NGC2423 No3 & 4545 $\pm$ 71 & 2.20 $\pm$ 0.20 & -0.08 $\pm$ 0.05 & 1.55 $\pm$ 0.07 & 1.18 $\pm$ 0.26 \\
NGC4349 No127 & 4445 $\pm$ 87 & 1.64 $\pm$ 0.23 & -0.25 $\pm$ 0.06 & 1.84 $\pm$ 0.08 & 1.37 $\pm$ 0.37 \\
nuOph & 4967 $\pm$ 61 & 2.70 $\pm$ 0.13 & 0.14 $\pm$ 0.05 & 1.61 $\pm$ 0.06 & 2.94 $\pm$ 0.08 \\

\end{longtable}
\tablefoot{$^{\ast\ast}$ For HD13189, the SO08 parameters are mentioned and adopted, even though the star is cooler than 5200\,K.}
\end{longtab}

\begin{longtab}
\begin{longtable}{lccccc}
\caption{\label{TabParMel} Stellar parameters derived with the HM07 line list. For stars with an asterisk, the microturbulences were derived with a calibration formula.}\\
\hline\hline
Name & T$_{eff}$ & $\log g_{spec}$ & [Fe/H] & $\xi$ & M$_{\ast}$ \\
 & (K) & (cm s$^{-2}$) & (dex) & (km s$^{-1}$)  & (M$_{\odot}$)  \\
\hline
\endfirsthead
\caption{continued.}\\
\hline\hline
Name & T$_{eff}$ & $\log g_{spec}$ & [Fe/H] & $\xi$ & M$_{\ast}$ \\
 & (K) & (cm s$^{-2}$) & (dex) & (km s$^{-1}$)  & (M$_{\odot}$)  \\
\hline
\endhead
\hline
\endfoot
11Com & 4811 $\pm$ 76 & 2.70 $\pm$ 0.14 & -0.29 $\pm$ 0.06 & 1.59 $\pm$ 0.10 & 2.09 $\pm$ 0.29 \\
18Del & 5090 $\pm$ 71 & 3.23 $\pm$ 0.11 & 0.07 $\pm$ 0.06 & 1.25 $\pm$ 0.09 & 2.11 $\pm$ 0.13 \\
24Sex$^{\ast}$ & 5061 $\pm$ 88 & 3.65 $\pm$ 0.15 & 0.12 $\pm$ 0.08 & 1.12 $\pm$ 0.20 & 1.86 $\pm$ 0.11 \\
75Cet$^{\ast}$ & 4853 $\pm$ 128 & 2.84 $\pm$ 0.23 & 0.15 $\pm$ 0.13 & 1.23 $\pm$ 0.20 & 1.88 $\pm$ 0.40 \\
7CMa$^{\ast}$ & 4790 $\pm$ 138 & 3.37 $\pm$ 0.28 & 0.28 $\pm$ 0.09 & 1.26 $\pm$ 0.20 & 1.30 $\pm$ 0.16 \\
81Cet & 4858 $\pm$ 97 & 2.63 $\pm$ 0.26 & 0.01 $\pm$ 0.09 & 1.56 $\pm$ 0.11 & 1.77 $\pm$ 0.39 \\
91Aqr & 4637 $\pm$ 62 & 2.50 $\pm$ 0.13 & -0.05 $\pm$ 0.05 & 1.57 $\pm$ 0.07 & 1.17 $\pm$ 0.16 \\
alfAri & 4614 $\pm$ 94 & 2.80 $\pm$ 0.21 & -0.05 $\pm$ 0.06 & 1.46 $\pm$ 0.10 & 1.58 $\pm$ 0.24 \\
BD+202457 & 4196 $\pm$ 103 & 1.59 $\pm$ 0.26 & -0.77 $\pm$ 0.05 & 1.90 $\pm$ 0.12 & 1.07 $\pm$ 0.22 \\
BD+20274 & 4254 $\pm$ 170 & 1.80 $\pm$ 0.46 & -0.36 $\pm$ 0.08 & 1.84 $\pm$ 0.16 & 1.17 $\pm$ 0.25 \\
BD+48738 & 4493 $\pm$ 154 & 2.41 $\pm$ 0.35 & -0.03 $\pm$ 0.10 & 1.53 $\pm$ 0.15 & 1.19 $\pm$ 0.25 \\
epsCrB & 4272 $\pm$ 220 & 1.81 $\pm$ 0.54 & -0.21 $\pm$ 0.11 & 1.73 $\pm$ 0.19 & 1.23 $\pm$ 0.27 \\
epsTau$^{\ast}$ & 4812 $\pm$ 178 & 2.54 $\pm$ 0.33 & 0.24 $\pm$ 0.16 & 1.25 $\pm$ 0.20 & 2.63 $\pm$ 0.22 \\
gam01Leo & 4373 $\pm$ 76 & 1.87 $\pm$ 0.18 & -0.41 $\pm$ 0.05 & 1.72 $\pm$ 0.08 & 1.41 $\pm$ 0.24 \\
gammaCephei$^{\ast}$ & 4932 $\pm$ 259 & 3.63 $\pm$ 0.48 & 0.31 $\pm$ 0.13 & 1.18 $\pm$ 0.20 & 1.30 $\pm$ 0.19 \\
HD100655 & 4869 $\pm$ 70 & 3.05 $\pm$ 0.14 & 0.13 $\pm$ 0.05 & 1.44 $\pm$ 0.08 & 2.08 $\pm$ 0.15 \\
HD102329 & 4751 $\pm$ 69 & 3.07 $\pm$ 0.15 & 0.11 $\pm$ 0.05 & 1.42 $\pm$ 0.08 & 1.29 $\pm$ 0.13 \\
HD102272 & 4790 $\pm$ 56 & 2.75 $\pm$ 0.10 & -0.36 $\pm$ 0.05 & 1.61 $\pm$ 0.09 & 1.09 $\pm$ 0.16 \\
HD104985 & 4750 $\pm$ 71 & 2.78 $\pm$ 0.13 & -0.26 $\pm$ 0.05 & 1.51 $\pm$ 0.10 & 1.01 $\pm$ 0.16 \\
HD106270 & 5611 $\pm$ 76 & 4.08 $\pm$ 0.08 & 0.09 $\pm$ 0.07 & 1.48 $\pm$ 0.19 & 1.33 $\pm$ 0.06 \\
HD108863 & 4906 $\pm$ 89 & 3.29 $\pm$ 0.16 & 0.14 $\pm$ 0.06 & 1.27 $\pm$ 0.11 & 1.42 $\pm$ 0.15 \\
HD110014 & 4430 $\pm$ 233 & 2.45 $\pm$ 0.54 & 0.20 $\pm$ 0.14 & 1.83 $\pm$ 0.24 & 1.82 $\pm$ 0.60 \\
HD116029 & 4846 $\pm$ 82 & 3.35 $\pm$ 0.16 & 0.13 $\pm$ 0.05 & 1.24 $\pm$ 0.09 & 1.38 $\pm$ 0.14 \\
HD11964 & 5375 $\pm$ 79 & 4.17 $\pm$ 0.12 & 0.16 $\pm$ 0.06 & 1.01 $\pm$ 0.12 & 1.08 $\pm$ 0.02 \\
HD11977 & 5048 $\pm$ 78 & 3.17 $\pm$ 0.11 & -0.08 $\pm$ 0.06 & 1.58 $\pm$ 0.10 & 2.32 $\pm$ 0.16 \\
HD122430 & 4264 $\pm$ 174 & 2.06 $\pm$ 0.46 & -0.11 $\pm$ 0.10 & 1.75 $\pm$ 0.18 & 1.26 $\pm$ 0.29 \\
HD13189 & 4228 $\pm$ 242 & 2.09 $\pm$ 0.61 & -0.52 $\pm$ 0.14 & 1.88 $\pm$ 0.30 & 1.08 $\pm$ 0.16 \\
HD148427$^{\ast}$ & 5024 $\pm$ 107 & 3.68 $\pm$ 0.18 & 0.14 $\pm$ 0.08 & 1.14 $\pm$ 0.20 & 1.37 $\pm$ 0.11 \\
HD1502 & 5002 $\pm$ 73 & 3.44 $\pm$ 0.12 & 0.01 $\pm$ 0.06 & 1.31 $\pm$ 0.10 & 1.36 $\pm$ 0.11 \\
HD156411 & 5892 $\pm$ 80 & 4.16 $\pm$ 0.07 & -0.06 $\pm$ 0.07 & 1.30 $\pm$ 0.31 & 1.27 $\pm$ 0.05 \\
HD159868 & 5514 $\pm$ 92 & 4.08 $\pm$ 0.08 & -0.10 $\pm$ 0.08 & 1.32 $\pm$ 0.23 & 1.16 $\pm$ 0.05 \\
HD167042$^{\ast}$ & 5061 $\pm$ 101 & 3.74 $\pm$ 0.17 & 0.20 $\pm$ 0.10 & 1.12 $\pm$ 0.20 & 1.68 $\pm$ 0.10 \\
HD1690 & 4157 $\pm$ 186 & 1.54 $\pm$ 0.48 & -0.27 $\pm$ 0.07 & 1.46 $\pm$ 0.13 & 1.20 $\pm$ 0.29 \\
HD171028 & 5697 $\pm$ 89 & 4.18 $\pm$ 0.08 & -0.44 $\pm$ 0.07 & 1.74 $\pm$ 0.66 & 0.89 $\pm$ 0.02 \\
HD175541 & 5093 $\pm$ 88 & 3.66 $\pm$ 0.13 & -0.09 $\pm$ 0.07 & 1.2 $\pm$ 0.12 & 1.20 $\pm$ 0.08 \\
HD177830$^{\ast}$ & 4881 $\pm$ 204 & 3.91 $\pm$ 0.38 & 0.39 $\pm$ 0.11 & 1.21 $\pm$ 0.20 & 1.22 $\pm$ 0.12 \\
HD180902$^{\ast}$ & 5001 $\pm$ 99 & 3.48 $\pm$ 0.18 & 0.08 $\pm$ 0.10 & 1.15 $\pm$ 0.20 & 1.41 $\pm$ 0.16 \\
HD181342$^{\ast}$ & 4930 $\pm$ 122 & 3.39 $\pm$ 0.22 & 0.24 $\pm$ 0.11 & 1.19 $\pm$ 0.20 & 1.49 $\pm$ 0.19 \\
HD18742 & 5007 $\pm$ 79 & 3.37 $\pm$ 0.13 & -0.14 $\pm$ 0.07 & 1.35 $\pm$ 0.12 & 1.28 $\pm$ 0.12 \\
HD192699 & 5126 $\pm$ 72 & 3.56 $\pm$ 0.11 & -0.20 $\pm$ 0.06 & 1.27 $\pm$ 0.13 & 1.48 $\pm$ 0.10 \\
HD200964 & 5078 $\pm$ 68 & 3.53 $\pm$ 0.10 & -0.04 $\pm$ 0.06 & 1.06 $\pm$ 0.10 & 1.67 $\pm$ 0.10 \\
HD206610$^{\ast}$ & 4830 $\pm$ 125 & 3.42 $\pm$ 0.23 & 0.14 $\pm$ 0.10 & 1.24 $\pm$ 0.20 & 1.20 $\pm$ 0.12 \\
HD210702 & 4996 $\pm$ 81 & 3.50 $\pm$ 0.14 & 0.06 $\pm$ 0.06 & 1.21 $\pm$ 0.10 & 1.63 $\pm$ 0.13 \\
HD212771 & 5080 $\pm$ 78 & 3.63 $\pm$ 0.13 & -0.11 $\pm$ 0.06 & 1.31 $\pm$ 0.11 & 1.22 $\pm$ 0.08 \\
HD27442$^{\ast}$ & 4768 $\pm$ 136 & 3.50 $\pm$ 0.31 & 0.21 $\pm$ 0.11 & 1.27 $\pm$ 0.20 & 1.19 $\pm$ 0.12 \\
HD28678 & 5088 $\pm$ 68 & 3.41 $\pm$ 0.10 & -0.09 $\pm$ 0.06 & 1.27 $\pm$ 0.10 & 1.40 $\pm$ 0.12 \\
HD30856 & 4964 $\pm$ 79 & 3.41 $\pm$ 0.13 & -0.11 $\pm$ 0.06 & 1.29 $\pm$ 0.10 & 1.30 $\pm$ 0.13 \\
HD33142 & 5029 $\pm$ 86 & 3.60 $\pm$ 0.14 & 0.08 $\pm$ 0.06 & 1.22 $\pm$ 0.10 & 1.56 $\pm$ 0.14 \\
HD38529 & 5547 $\pm$ 52 & 3.69 $\pm$ 0.12 & 0.28 $\pm$ 0.05 & 1.29 $\pm$ 0.08 & 1.37 $\pm$ 0.02 \\
HD38801 & 5241 $\pm$ 90 & 3.98 $\pm$ 0.14 & 0.27 $\pm$ 0.07 & 1.22 $\pm$ 0.13 & 1.24 $\pm$ 0.07 \\
HD4313 & 4993 $\pm$ 86 & 3.49 $\pm$ 0.14 & 0.08 $\pm$ 0.07 & 1.33 $\pm$ 0.11 & 1.35 $\pm$ 0.11 \\
HD47536 & 4500 $\pm$ 194 & 2.54 $\pm$ 0.39 & -0.58 $\pm$ 0.10 & 1.83 $\pm$ 0.19 & 1.18 $\pm$ 0.25 \\
HD48265 & 5683 $\pm$ 81 & 4.07 $\pm$ 0.08 & 0.32 $\pm$ 0.07 & 1.33 $\pm$ 0.16 & 1.21 $\pm$ 0.04 \\
HD5319$^{\ast}$ & 4886 $\pm$ 104 & 3.43 $\pm$ 0.17 & 0.04 $\pm$ 0.10 & 1.21 $\pm$ 0.20 & 1.24 $\pm$ 0.14 \\
HD5608$^{\ast}$ & 4874 $\pm$ 122 & 3.34 $\pm$ 0.23 & 0.16 $\pm$ 0.10 & 1.21 $\pm$ 0.20 & 1.41 $\pm$ 0.19 \\
HD5891 & 4827 $\pm$ 80 & 2.84 $\pm$ 0.13 & -0.34 $\pm$ 0.07 & 1.75 $\pm$ 0.13 & 1.12 $\pm$ 0.16 \\
HD59686$^{\ast}$ & 4592 $\pm$ 228 & 2.59 $\pm$ 0.49 & 0.28 $\pm$ 0.14 & 1.36 $\pm$ 0.20 & 2.04 $\pm$ 0.46 \\
HD62509 & 4992 $\pm$ 235 & 3.46 $\pm$ 0.39 & 0.11 $\pm$ 0.17 & 1.61 $\pm$ 0.28 & 1.61 $\pm$ 0.54 \\
HD66141 & 4260 $\pm$ 117 & 1.93 $\pm$ 0.31 & -0.40 $\pm$ 0.06 & 1.65 $\pm$ 0.12 & 1.06 $\pm$ 0.12 \\
HD73534$^{\ast}$ & 5003 $\pm$ 129 & 4.06 $\pm$ 0.25 & 0.26 $\pm$ 0.08 & 1.15 $\pm$ 0.20 & 1.20 $\pm$ 0.07 \\
HD88133 & 5473 $\pm$ 22 & 3.83 $\pm$ 0.22 & 0.30 $\pm$ 0.02 & 1.17 $\pm$ 0.03 & 1.18 $\pm$ 0.06 \\
HD89744 & 6095 $\pm$ 97 & 3.77 $\pm$ 0.13 & 0.09 $\pm$ 0.08 & 1.66 $\pm$ 0.28 & 1.44 $\pm$ 0.05 \\
HD95089$^{\ast}$ & 4956 $\pm$ 90 & 3.51 $\pm$ 0.15 & 0.12 $\pm$ 0.09 & 1.17 $\pm$ 0.20 & 1.30 $\pm$ 0.11 \\
HD96063 & 5142 $\pm$ 62 & 3.74 $\pm$ 0.09 & -0.10 $\pm$ 0.05 & 1.12 $\pm$ 0.10 & 1.16 $\pm$ 0.07 \\
HD98219 & 4970 $\pm$ 94 & 3.65 $\pm$ 0.16 & 0.08 $\pm$ 0.07 & 1.17 $\pm$ 0.12 & 1.47 $\pm$ 0.16 \\
HIP75458 & 4950 $\pm$ 202 & 3.92 $\pm$ 0.36 & -0.03 $\pm$ 0.11 & 2.92 $\pm$ 0.62 & 1.57 $\pm$ 0.42 \\
kappaCrB & 4853 $\pm$ 132 & 3.41 $\pm$ 0.25 & 0.12 $\pm$ 0.08 & 1.27 $\pm$ 0.13 & 1.32 $\pm$ 0.17 \\
ksiAql & 4719 $\pm$ 80 & 2.83 $\pm$ 0.16 & -0.18 $\pm$ 0.06 & 1.58 $\pm$ 0.09 & 0.97 $\pm$ 0.17 \\
NGC2423 No3 & 4337 $\pm$ 144 & 1.69 $\pm$ 0.36 & -0.11 $\pm$ 0.07 & 1.36 $\pm$ 0.09 & 1.16 $\pm$ 0.34 \\
NGC4349 No127 & 4258 $\pm$ 160 & 1.52 $\pm$ 0.43 & -0.22 $\pm$ 0.07 & 1.80 $\pm$ 0.11 & 1.27 $\pm$ 0.30 \\
nuOph & 4919 $\pm$ 105 & 2.88 $\pm$ 0.18 & 0.14 $\pm$ 0.08 & 1.72 $\pm$ 0.12 & 2.91 $\pm$ 0.11 \\

\end{longtable}
\end{longtab}

With the TS13 line list, no solution could be found for \object{HD13189}. This star was observed with the SARG spectrograph at the TNG telescope in 2002. The spectrum was very noisy and the spectral lines were hard to measure. We calculated all EWs by hand, using the TS13 line list, but could not converge to a viable solution with MOOG. The results from the SO08 line list are thus used for this star. For \object{HIP75458}, the EWs were also measured by hand, using the TS13 line list.

Using the smaller HM07 line list did not always allow for good microturbulence determinations because of the small EW interval of the measured \ion{Fe}{i} lines. This is a disadvantage of using small line lists. For 17 stars, the microturbulence was thus derived with the formula

\[
\xi_t = 3.7 - 5.1\cdot10^{-4} T_{eff}.
\]

This formula was empirically derived by HM07 based on their results, using their small line list. These 17 stars are marked with an asterisk in Table \ref{TabParMel}. We ran a test for 10 stars, where we derived the atmospheric parameters both with the standard method and with this formula for the microturbulence derivation. We find that all parameters compare well, within errorbars. Especially for metallicity, the most important parameter in this work, the agreement is striking. The standard deviation for the differences for these stars are 54\,K, 0.19\,dex, 0.04\,dex, and 0.20\,km\,s$^{-1}$ for effective temperature, surface gravity, metallicity and microturbulence, respectively, while the mean error of these parameters are 94\,K, 0.16\,dex, 0.08\,dex, and 0.15\,km\,s$^{-1}$. This test shows that the use of the above-mentioned formula does not compromise the uniformity of the analysis.

When we used the HM07 line list for giant stars, there were also four stars where we calculated the EWs by hand: \object{HD104985}, \object{HD13189}, \object{HD177830}, and \object{HIP75458}. The stars were all observed with the SARG spectrograph at the TNG telescope. The spectra were not good enough to calculate the EWs automatically with ARES.

\begin{figure*}[t!]
\begin{center}
\includegraphics[width=6.5cm]{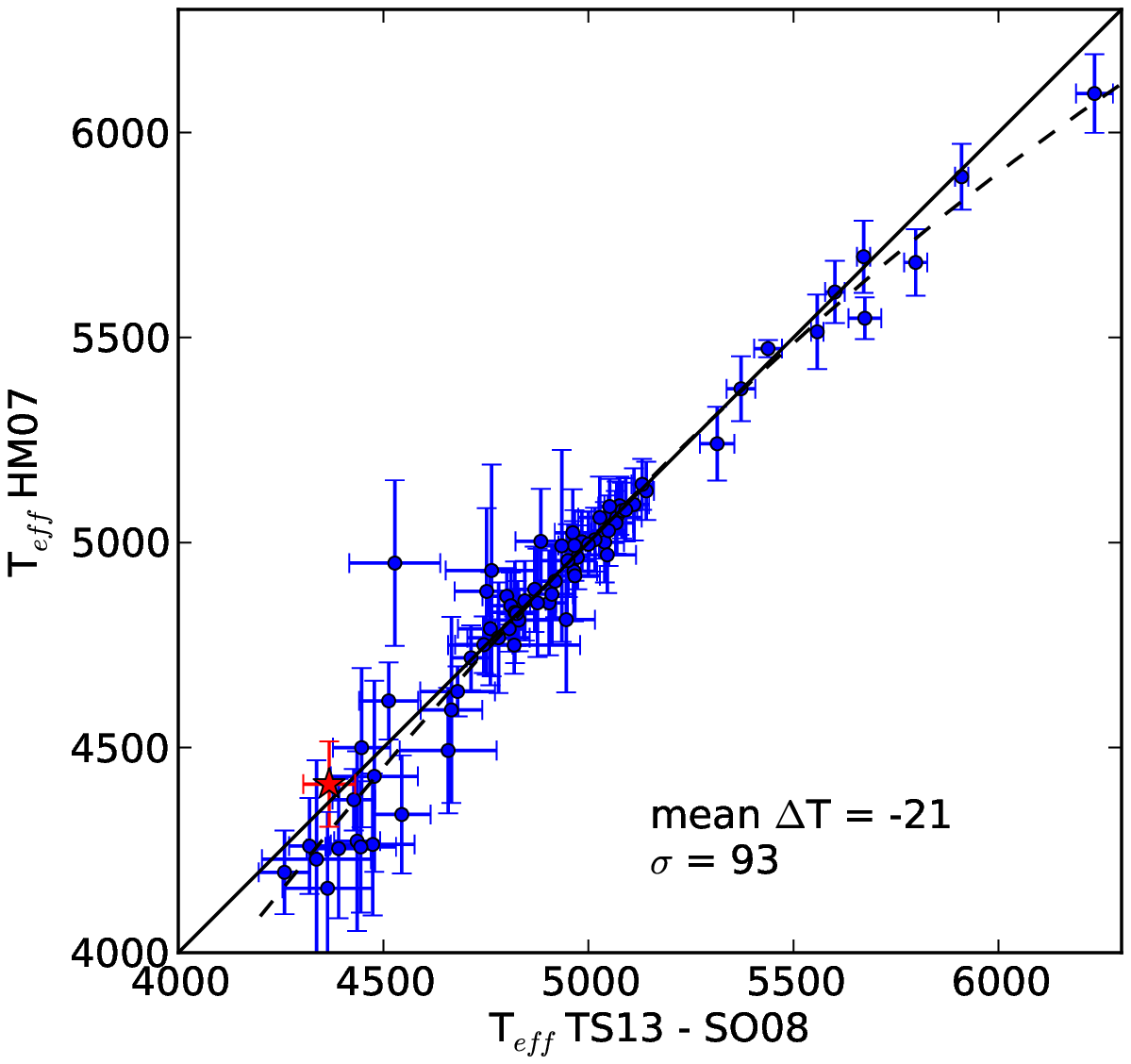}
\includegraphics[width=6.5cm]{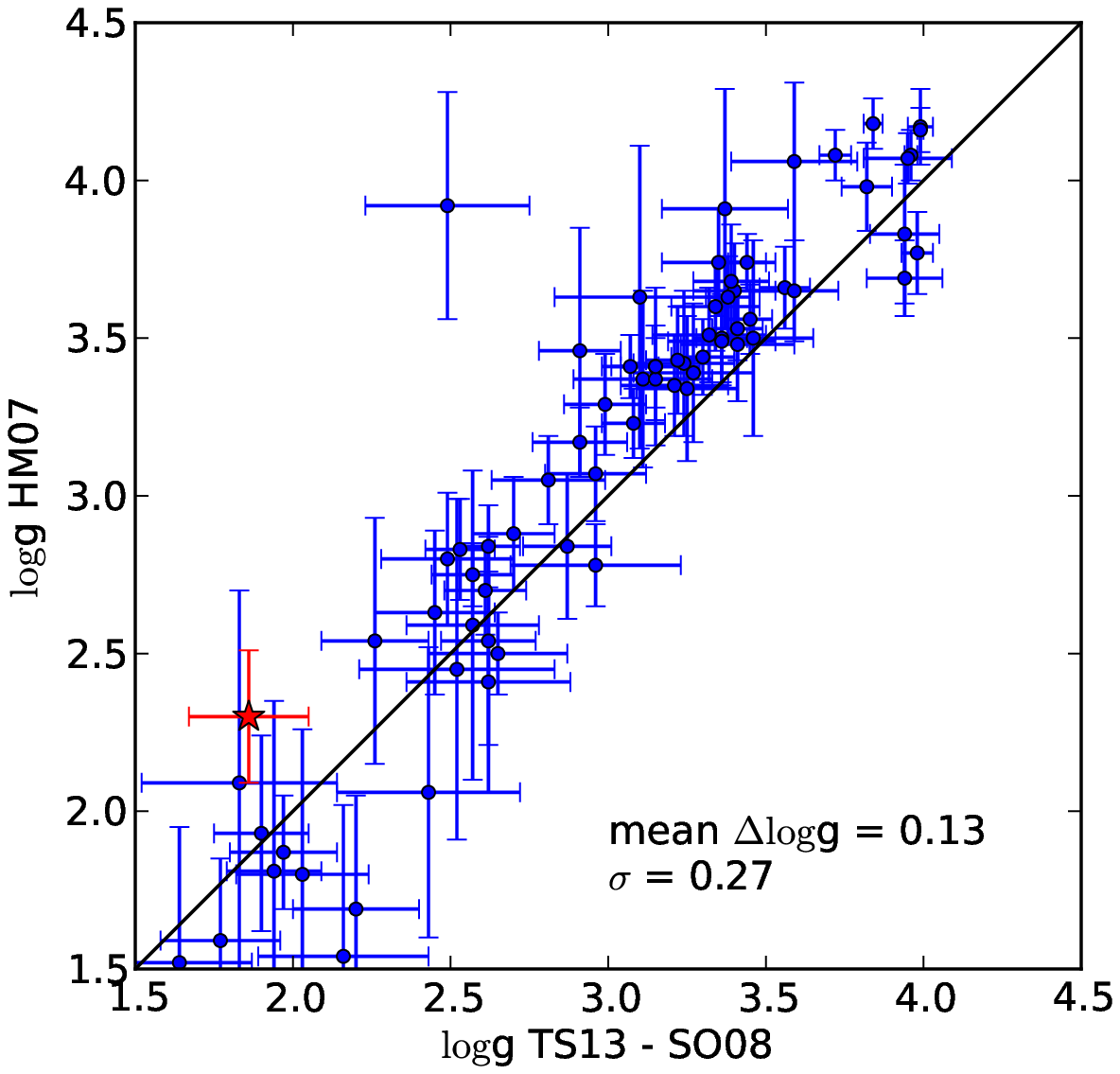}\\
\includegraphics[width=6.5cm]{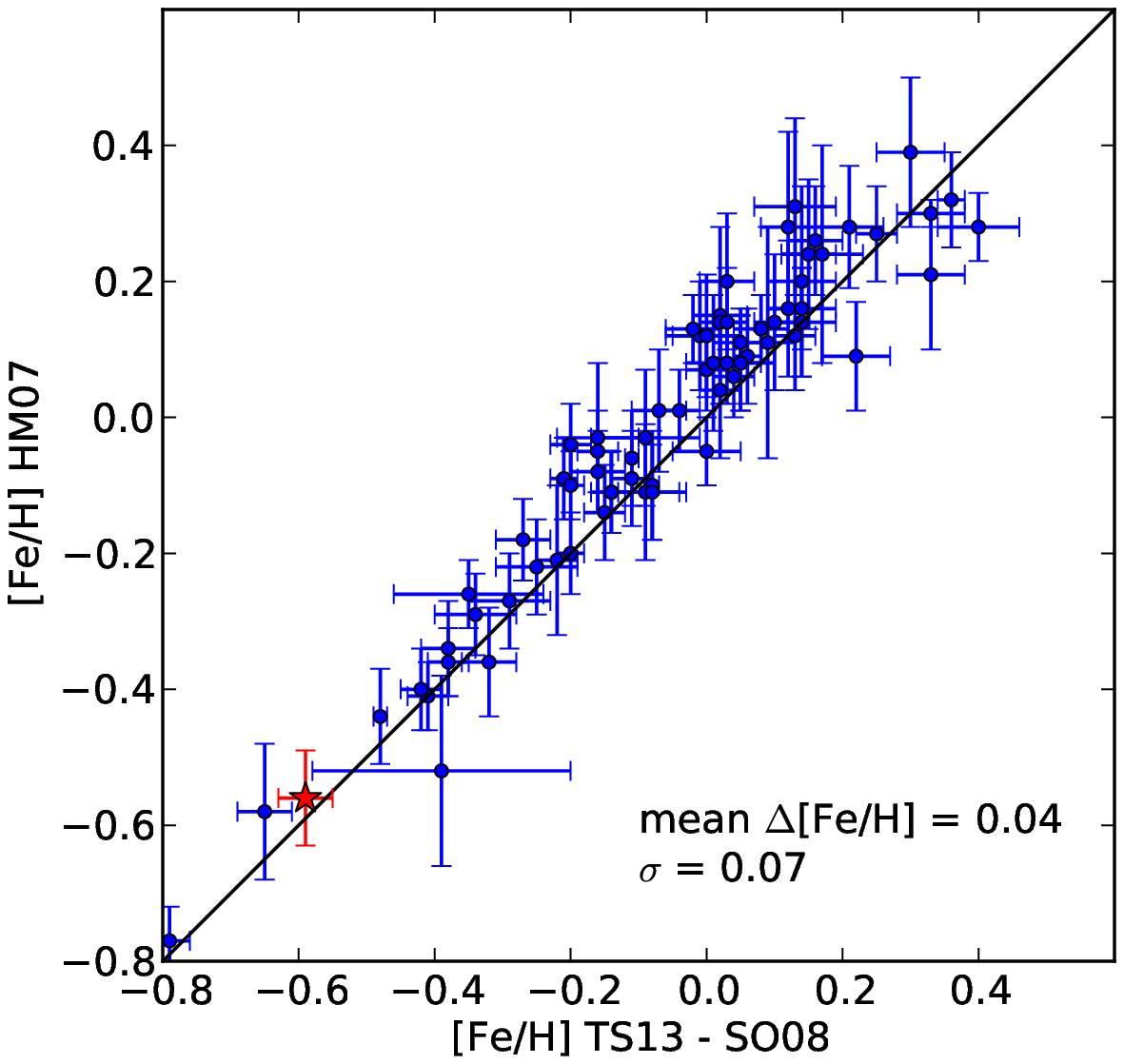}
\includegraphics[width=6.5cm]{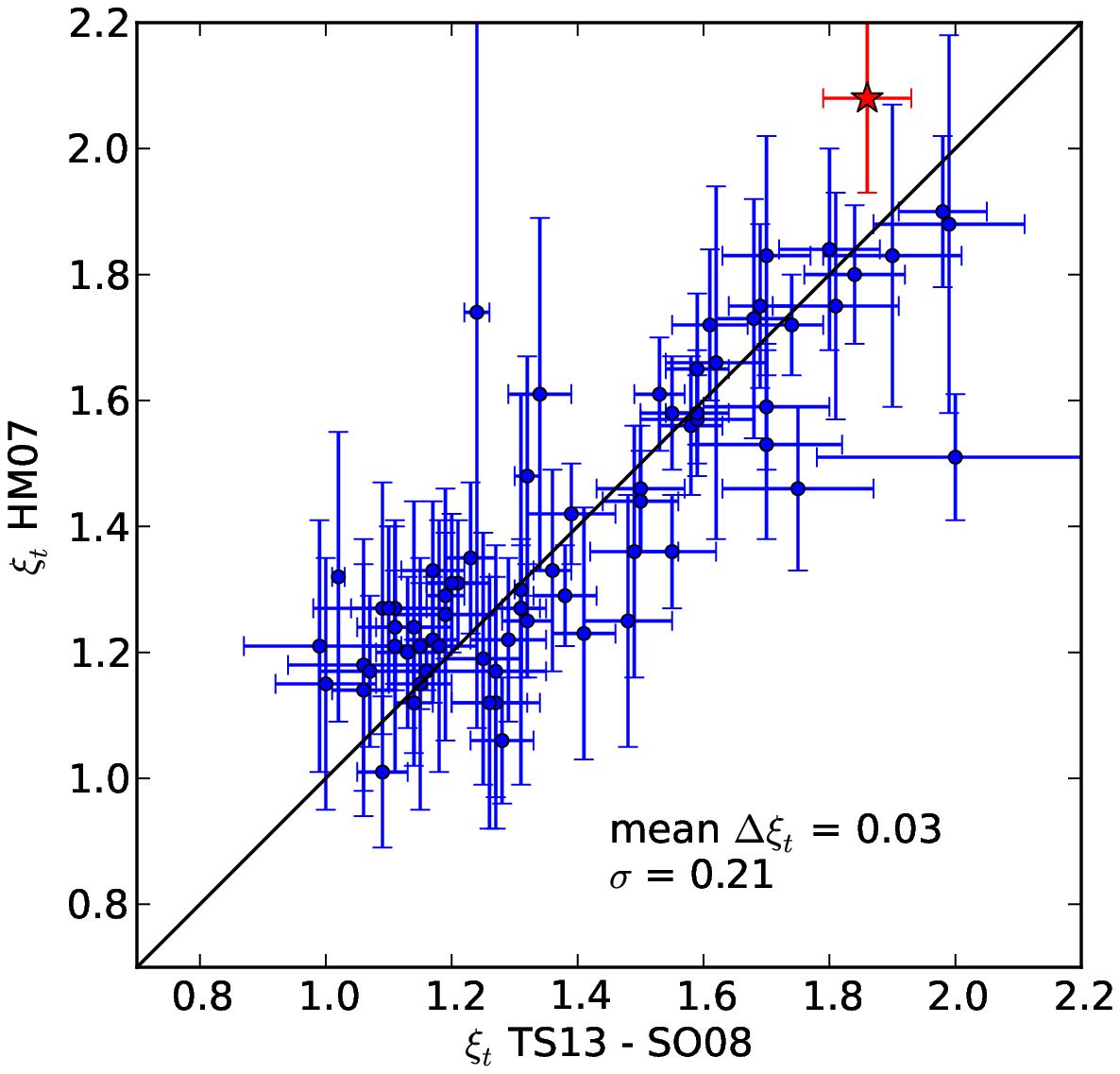}
\caption{Comparisons of the spectroscopic results from the TS13 - SO08 line list versus the HM07 line list for effective temperature, surface gravity, metallicity, and microturbulence. The measurements for the reference star Arcturus are overplotted with a star symbol. The dashed line in the top-left panel represents a second degree polynomial fit.}
\label{FigComp}
\end{center}
\end{figure*}

In Fig. \ref{FigComp}, we compare the atmospheric parameters derived with the different line lists. All differences presented in this work are calculated as y-axis - x-axis, as they are presented in Fig. \ref{FigComp}. The comparisons with the cool SO08 results are presented in Appendix \ref{ApSou}.

\subsection{Effective temperature}

The TS13 - SO08 line list set and the HM07 line list provide comparable temperatures. A mean difference of $-21$\,K is found between the results of HM07 and TS13 - SO08. The temperatures differ more for the coolest stars. If we only consider the temperatures higher than 4700\,K, there is a mean difference of only $-4$\,K. The values from the TS13 - SO08 line list set are preferred thanks to the good comparison with the HM07 line list, designed for giant stars, and the smaller errorbars (because the line lists of TS13 - SO08 have more lines than the HM07 line list). We fitted a second degree polynomial to the data, 

\[
T_{eff,HM07} = -0.00014 \cdot T_{eff,TS13}^2 + 2.39 \cdot T_{eff,TS13} - 3556,
\]

\noindent so one can correct for the slight offset amongst the coolest stars, if judged necessary. Our temperatures are not corrected with this polynomial.

\subsection{Surface gravity and microturbulence}

The determined surface gravities with the two line list sets are comparable with a mean difference of $0.13$\,dex for  HM07 versus TS13 - SO08 (see top right panel in Fig. \ref{FigComp}). Gravities determined with the HM07 line list are slightly higher than the other $\log$g values, but every value lies within the errorbars.

Microturbulences also compare very well with each other (bottom right panel in Fig. \ref{FigComp}) with a mean difference of $0.03$\,km/s. These results compare remarkably well. Thanks to these very good comparisons, the values from the TS13 - SO08 line list set were again preferred as a reference.

\subsection{Metallicity}\label{CompMet}

\citet{Hek07} found comparable metallicity results in their sample with literature data. They also compared their results with the homogeneous analysis of \citet{Luc07} for giant stars in the local region; HM07 find slightly lower metallicities than \citet{Luc07} with a mean difference of $-0.05$\,dex. They claim the differences are due to the use of different model atmospheres (Kurucz versus MARCS). We have derived parameters for our reference star Arcturus by using the spherical MARCS models and the TS13 line list. We find that the metallicities derived with the Kurucz models are marginally lower than the one derived with the MARCS model (see Table \ref{TabArc}).

We find slightly higher metallicities by using the HM07 line list with respect to using the TS13 - SO08 line list set with a mean differences of $0.04$\,dex. The results of the TS13 - SO08 line list set are thus even more different from the results of \citet{Luc07}. This offset in metallicities cannot be explained by different model atmospheres since both this work as HM07 use the Kurucz models. Furthermore, Fig. \ref{FigComp} shows that this offset is constant and does not vary with metallicity. It has to be noted that the offset is smaller than the spread. So, although the values from the HM07 line list are slightly higher, all values are comparable within errorbars. To guarantee uniformity, we adopt the metallicities derived with the TS13 - SO08 line list set. 

\subsection{Stellar masses}

\begin{figure}[t!]
\begin{center}
\includegraphics[width=6.7cm]{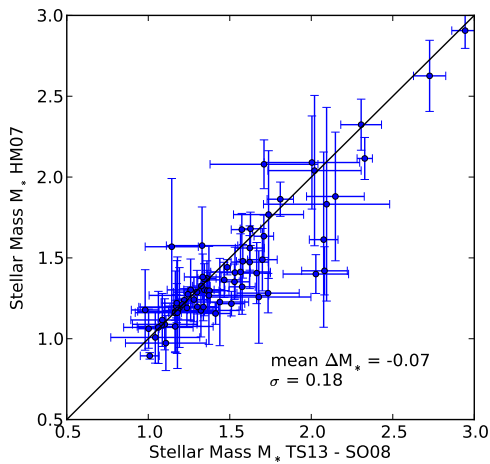}
\caption{Comparison of stellar masses derived with the line lists of TS13 and HM07.}
\label{FigMassMass}
\end{center}
\end{figure}

Stellar masses were calculated for the parameters derived with the TS13 - SO08 line list set and for the parameters derived with the HM07 line list. The results are listed in Tables \ref{TabParMar} and \ref{TabParMel}. A comparison between the values can be seen in Fig. \ref{FigMassMass}. The values agree within errorbars. They show a mean difference of $-0.07$\,M$_{\odot}$ between the values of HM07 and TS13 - SO08. We prefer to use the values derived with the TS13 - SO08 line list set for further analysis, as we do for the atmospheric parameters.

\section{Comparison with the literature}\label{Lit}

We used six different works \citep{Hek07,Das06,Das11,Val05,Gon10,Tak08} to compile a list of literature data for 31 stars in our sample (see Table \ref{TabLit}). These works all use spectroscopic methods to obtain the stellar parameters. We compared our stellar parameters (derived with the TS13 line list for the stars cooler than 5200\,K and the SO08 line list for the stars hotter than this value) with these literature values. 

\begin{figure*}[t!]
\begin{center}
\includegraphics[width=5.5cm]{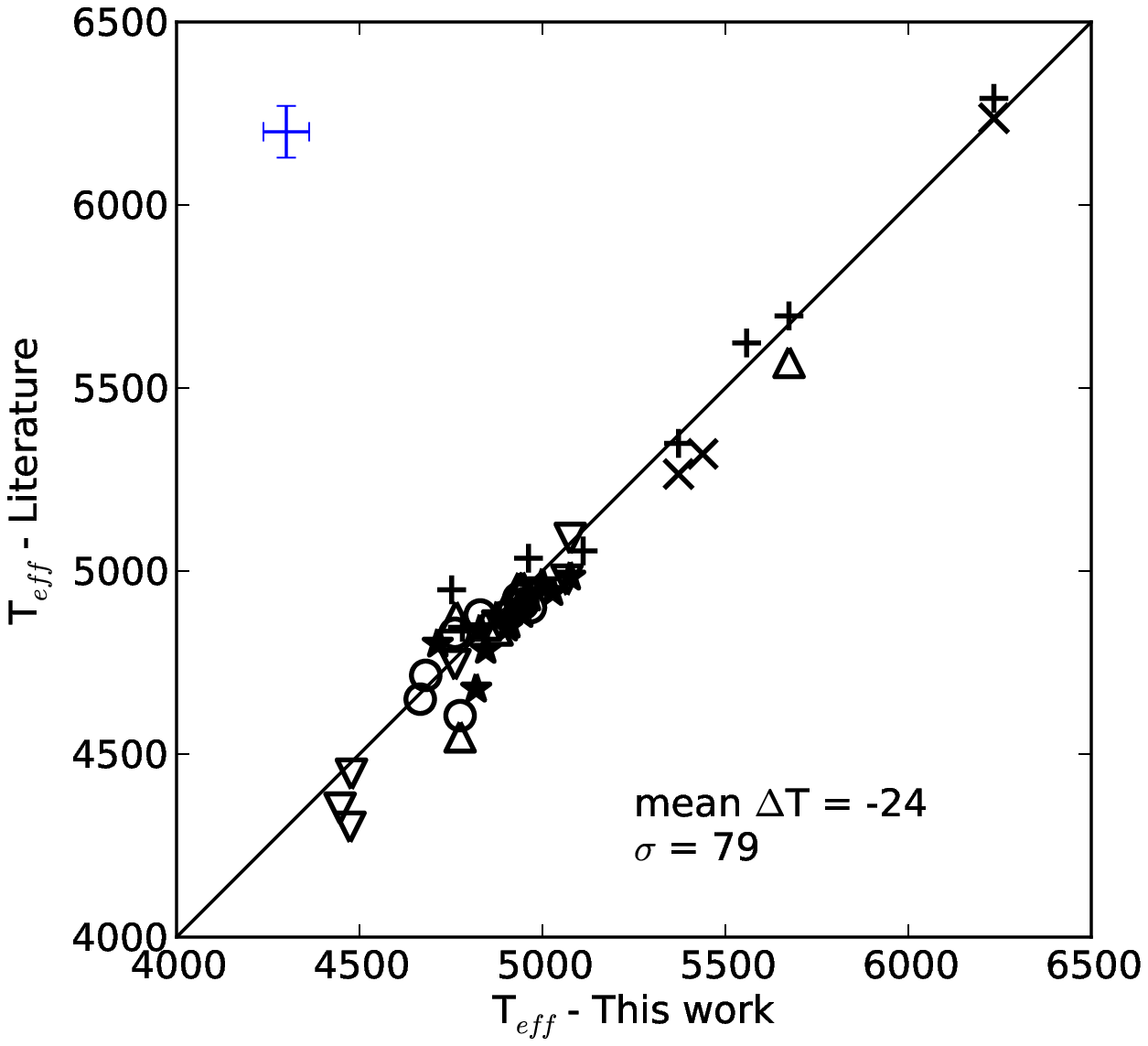}
\includegraphics[width=5.5cm]{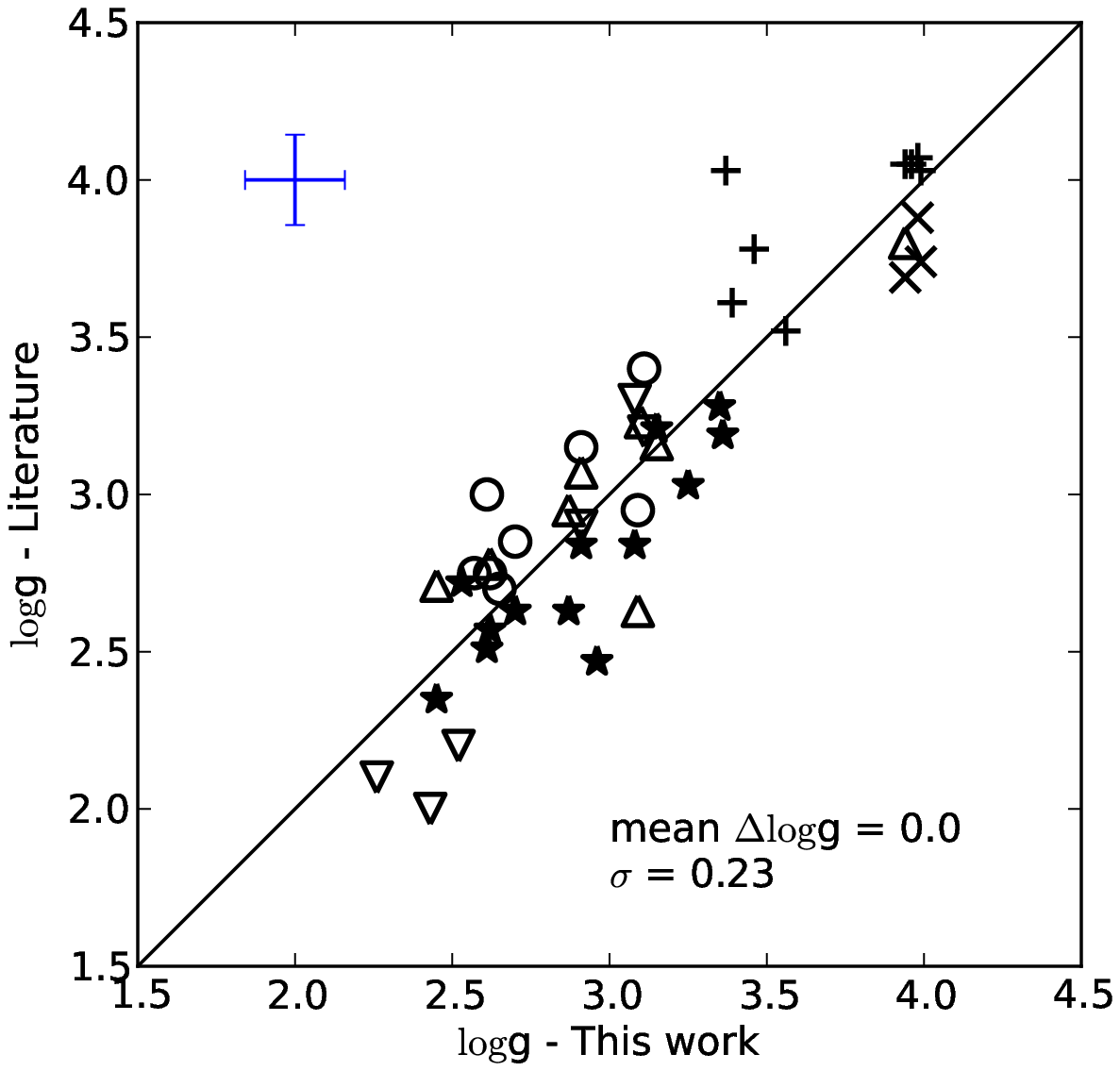}
\includegraphics[width=5.5cm]{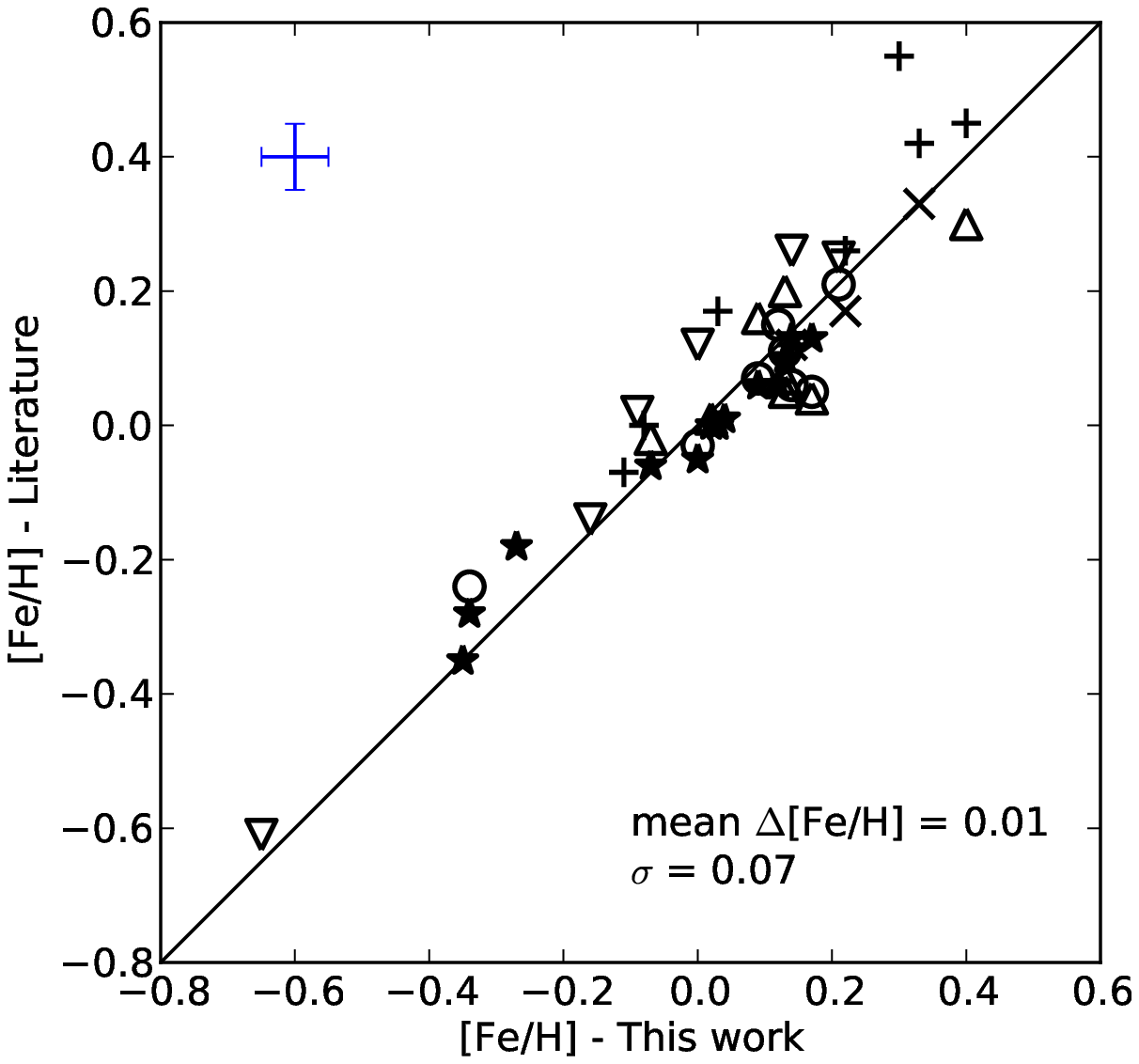}
\caption{Comparisons of the adopted spectroscopic results in this work with literature data for effective temperature, surface gravity and metallicity. Different symbols denote different references: circles for \citet{Hek07}, downward and upward triangles for \citet{Das06,Das11}, stars for \citet{Tak08}, + for \citet{Val05} and x for \citet{Gon10}.}
\label{FigLit}
\end{center}
\end{figure*}

The result is plotted in Fig. \ref{FigLit} where different symbols denote different references. The mean differences (defined as literature - this work) for temperature, surface gravity, and metallicity are -24\,K, 0.00\,dex, and 0.01\,dex. Given the spread in the data, we can say that the differences are essentially zero. This reinforces our choice for adopting the results derived from the TS13 - SO08 line list set.

Of particular relevance for the analysis in this paper is the comparison with the data from \citet{Das06} and \citet{Tak08} (see Sect. \ref{Tak}). For their values only, the mean differences with our values are -50\,K, -0.11\,dex, and 0.01\,dex for temperature, surface gravity, and metallicity, respectively.

\begin{table}
\caption{Atmospheric parameters from the literature for 31 different stars in our sample.}
\label{TabLit}
\centering
\begin{tabular}{ccccc}
\hline\hline
Name & T$_{eff}$ & $\log g_{spec}$ & [Fe/H] & Reference \\
 & (K) & (cm s$^{-2}$) & (dex) &  \\
\hline
11Com & 4880 & 3.00 & -0.24 & 1 \\
 & 4841 & 2.51 & -0.28 & 2 \\
18Del & 5089 & 3.30 & 0.12 & 3 \\
 & 4985 & 2.84 & -0.05 & 2 \\
7CMa & 4744 & 3.20 & 0.25 & 3 \\
 & 4830 & 3.40 & 0.21 & 1 \\
75Cet & 4906 & 2.95 & 0.01 & 4 \\
 & 4846 & 2.63 & 0.00 & 2 \\
81Cet & 4785 & 2.35 & -0.06 & 2 \\
 & 4853 & 2.71 & -0.02 & 4 \\
91Aqr & 4715 & 2.70 & -0.03 & 1 \\
epsTau & 4910 & 2.75 & 0.05 & 1 \\
 & 4956 & 2.78 & 0.04 & 4 \\
 & 4883 & 2.57 & 0.13 & 2 \\
gammaCephei & 4875 & 3.23 & 0.05 & 4 \\
HD104985 & 4679 & 2.47 & -0.35 & 2 \\
HD110014 & 4445 & 2.20 & 0.26 & 3 \\
HD11964 & 5349 & 4.03 & 0.12 & 5 \\
 & 5265 & 3.74 & 0.12 & 6 \\
HD11977 & 4975 & 2.90 & -0.14 & 3 \\
HD122430 & 4300 & 2.00 & 0.02 & 3 \\
HD148427 & 5035 & 3.61 & 0.17 & 5 \\
HD159868 & 5623 & 4.05 & 0.00 & 5 \\
HD167042 & 4943 & 3.28 & 0.00 & 2 \\
HD175541 & 5055 & 3.52 & -0.07 & 5 \\
HD177830 & 4949 & 4.03 & 0.55 & 5 \\
HD210702 & 4967 & 3.19 & 0.01 & 2 \\
HD27442 & 4846 & 3.78 & 0.42 & 5 \\
HD38529 & 5697 & 4.05 & 0.45 & 5 \\
 & 5570 & 3.80 & 0.30 & 4 \\
HD47536 & 4352 & 2.10 & -0.61 & 3 \\
HD5608 & 4854 & 3.03 & 0.06 & 2 \\
HD59686 & 4650 & 2.75 & 0.15 & 1 \\
HD62509 & 4925 & 3.15 & 0.07 & 1 \\
 & 4955 & 3.07 & 0.16 & 4 \\
 & 4904 & 2.84 & 0.06 & 2 \\
HD88133 & 5320 & 3.69 & 0.33 & 6 \\
HD89744 & 6291 & 4.07 & 0.26 & 5 \\
 & 6237 & 3.88 & 0.17 & 6 \\
HIP75458 & 4605 & 2.95 & 0.11 & 1 \\
 & 4547 & 2.63 & 0.07 & 4 \\
kappaCrB & 4839 & 3.16 & 0.20 & 4 \\
 & 4877 & 3.21 & 0.10 & 2 \\
ksiAql & 4802 & 2.72 & -0.18 & 2 \\
nuOph & 4900 & 2.85 & 0.06 & 1 \\
 & 4928 & 2.63 & 0.13 & 2 \\

\hline
\end{tabular}
\tablebib{(1)~\citet{Hek07};
(2) \citet{Tak08}; (3) \citet{Das06}; (4) \citet{Das11};
(5) \citet{Val05}; (6) \citet{Gon10}.
}
\end{table}

\section{Planet frequency}\label{Freq}

All evolved stars in this sample are orbited by at least one giant planet (with a mass between $0.1$ and $25$ M$_{Jup}$). For main-sequence FGK dwarfs, it has been shown that stars hosting a giant planet are more metal-rich than average field dwarfs \citep[e.g. ][]{San01,San04,Fis05}. There have been studies to check whether this metallicity enhancement can also be found in planet-hosting giants \citep{Hek07,Pas07,Tak08,Ghe10,Zie10,Mal13} with contradictory results.

\citet{Pas07} suggest that evolved planet-hosting stars are 0.2-0.3\,dex more metal-poor than main-sequence planet-hosting stars. Interestingly this is the same difference as between planet hosting and field dwarfs. \citet{Tak08} used a sample of late-G giants to show that there is no metallicity enhancement for giant stars hosting a giant planet. This result was confirmed by the preliminary results of \citet{Zie10} and the recent results of \citet{Mal13}. \citet{Hek07} and \citet{Ghe10} on the other hand do find a metallicity enhancement of about 0.13-0.21\,dex, comparable with what has been found in dwarf stars.

In the following sections we will use the results obtained using the TS13 - SO08 line list set to study the giant planet frequency. We will first explore the differences between dwarf and giant stars and discuss a bias in the planet search samples of giant stars. Then, we will compare planet-hosting giant stars with non-planet-hosting giant stars to see if there indeed is a metallicity enhancement for planet-hosting giants. For every comparison, we will also mention the results from the parameters derived using the HM07 line list.

\subsection{Giants versus dwarfs}

To tackle this problem, we first compared the metallicity distribution for our evolved planet hosts with that for dwarf planet hosts. For this purpose, we used the CORALIE+HARPS dwarf sample \citep[for details see ][ and references therein]{Sou11b,ME13}. For a full description of these surveys, see \citet{Mayor11}. The metallicities from this sample were derived using either the method described in this work, or the cross correlation function (CCF) calibration when no high-resolution spectra were available. The CCF calibration was derived by \citet{San04} who also showed that metallicities calculated with this calibration are compatible with the metallicities derived by the method used in this work. Furthermore, the dispersion of the calibration is only 0.06\,dex.

\begin{figure*}[t!]
\begin{center}
\includegraphics[width=5.7cm]{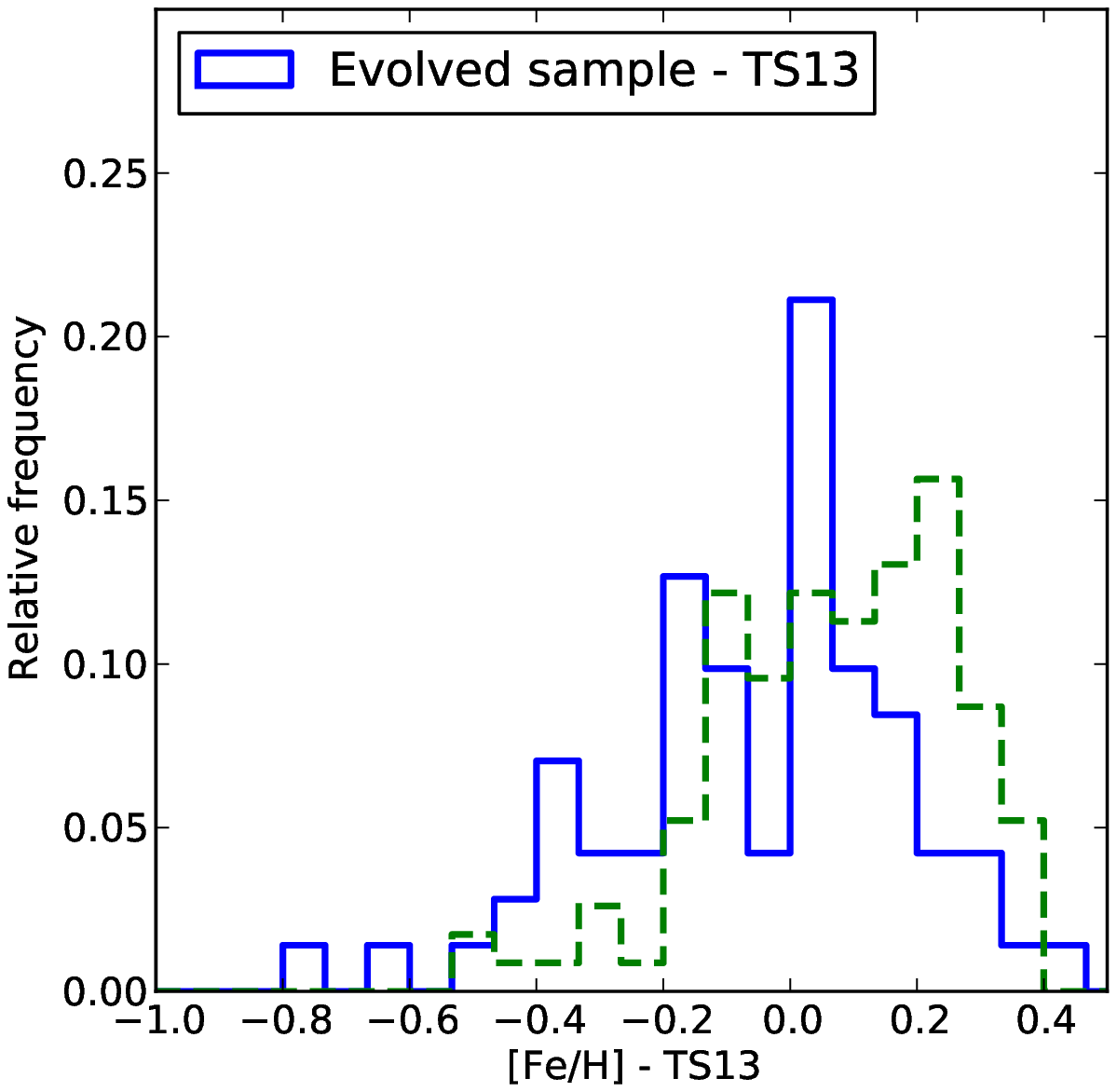}
\includegraphics[width=5.7cm]{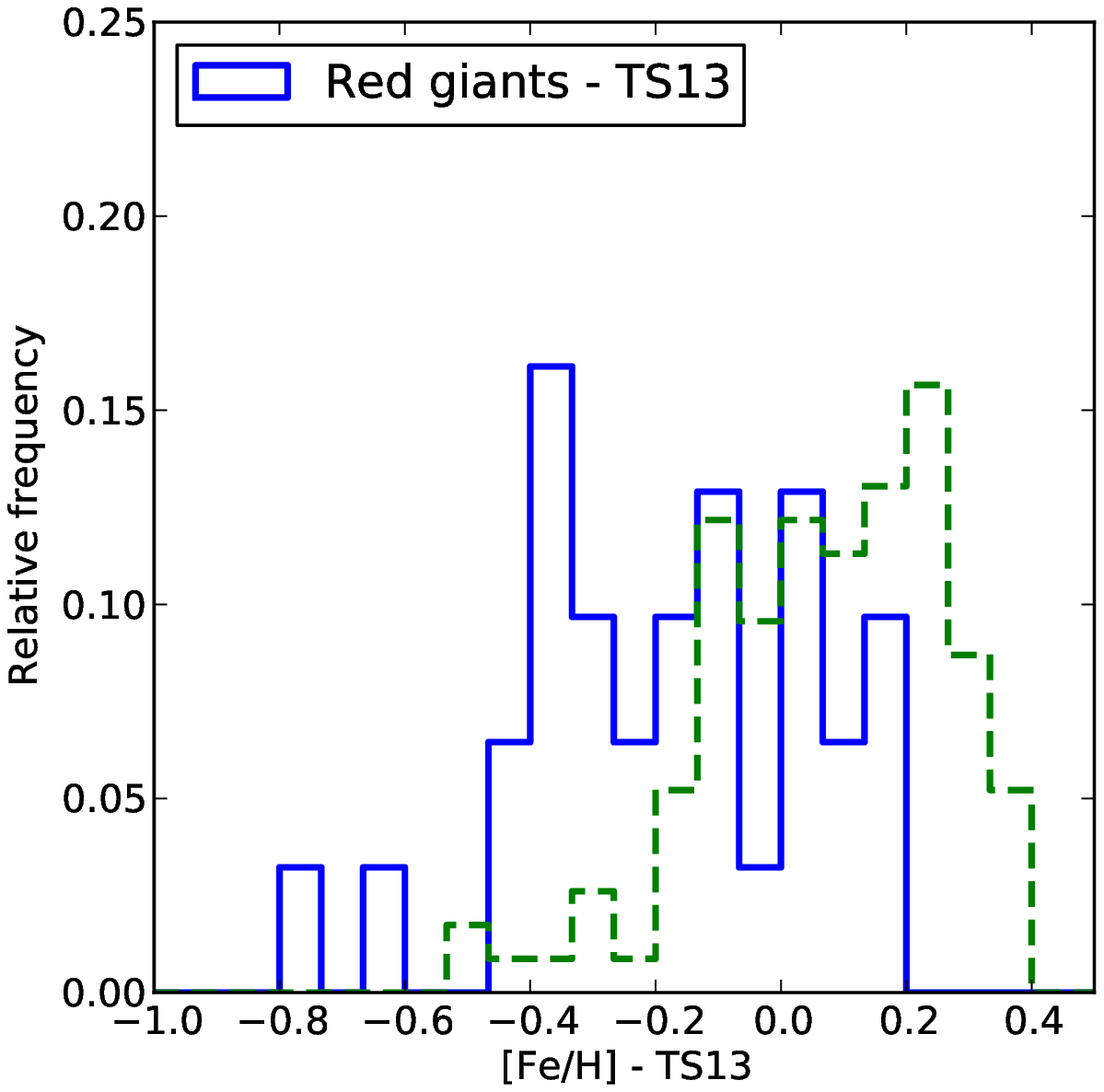}
\includegraphics[width=5.7cm]{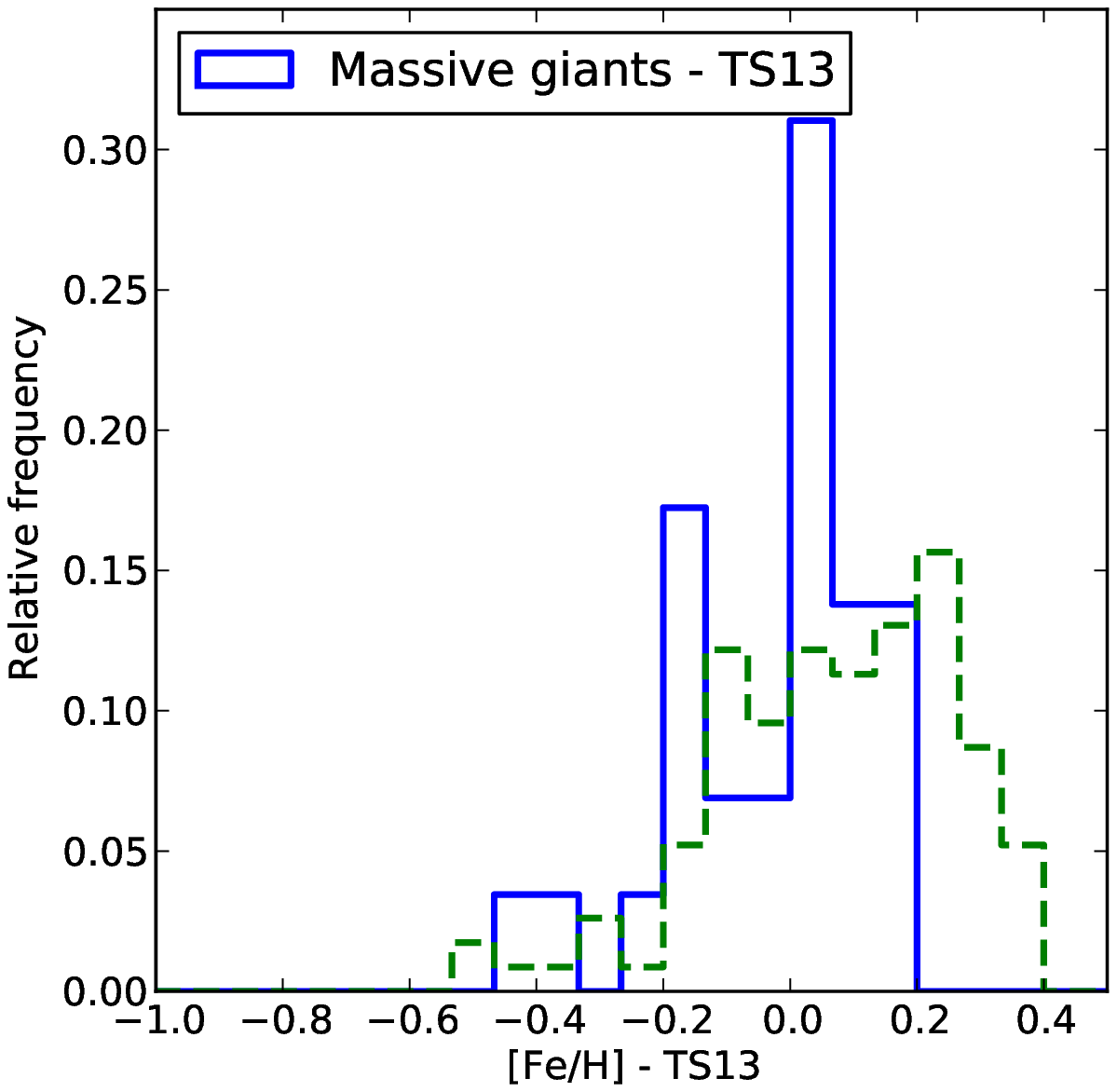}\\
\includegraphics[width=5.7cm]{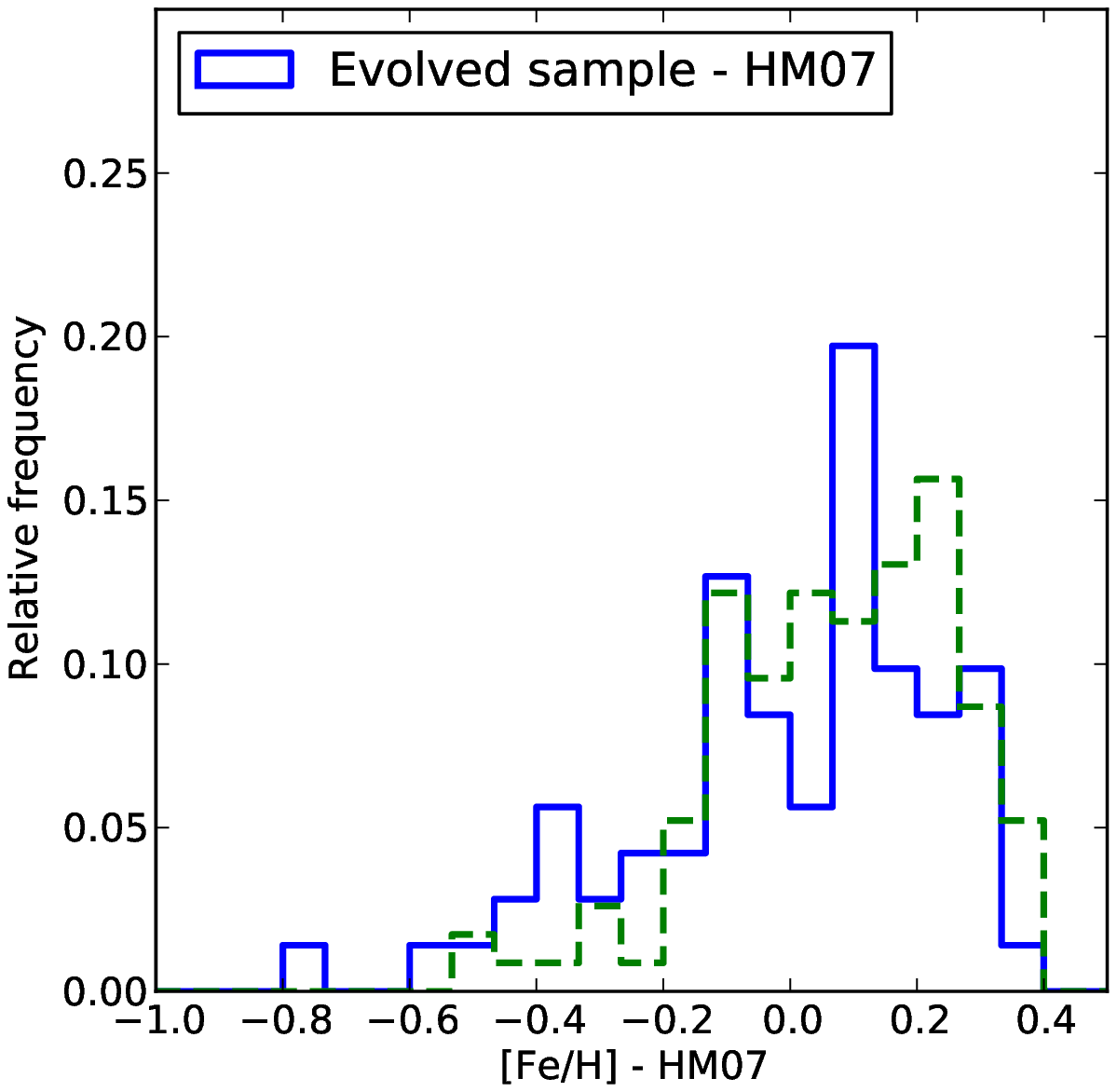}
\includegraphics[width=5.7cm]{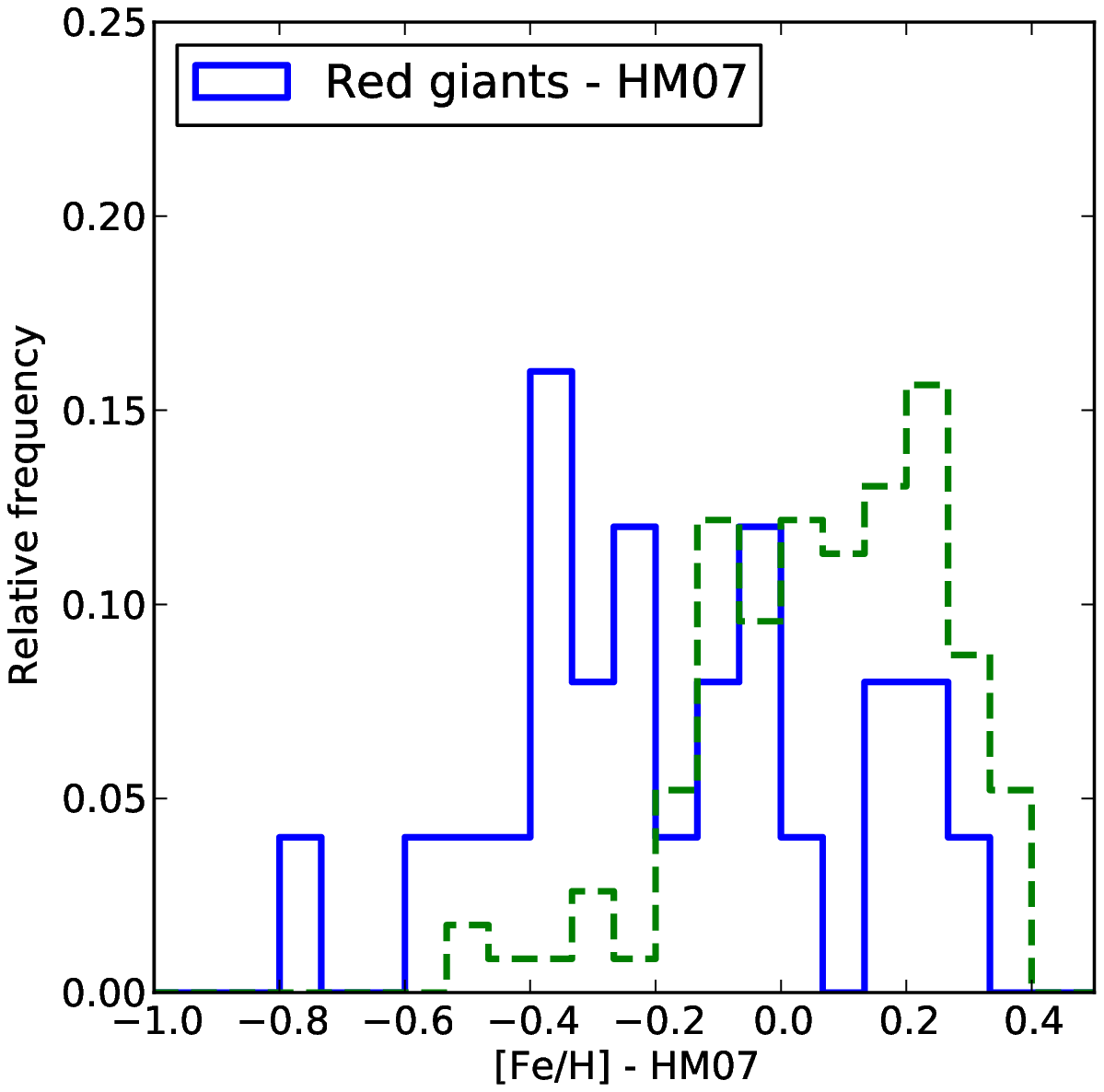}
\includegraphics[width=5.7cm]{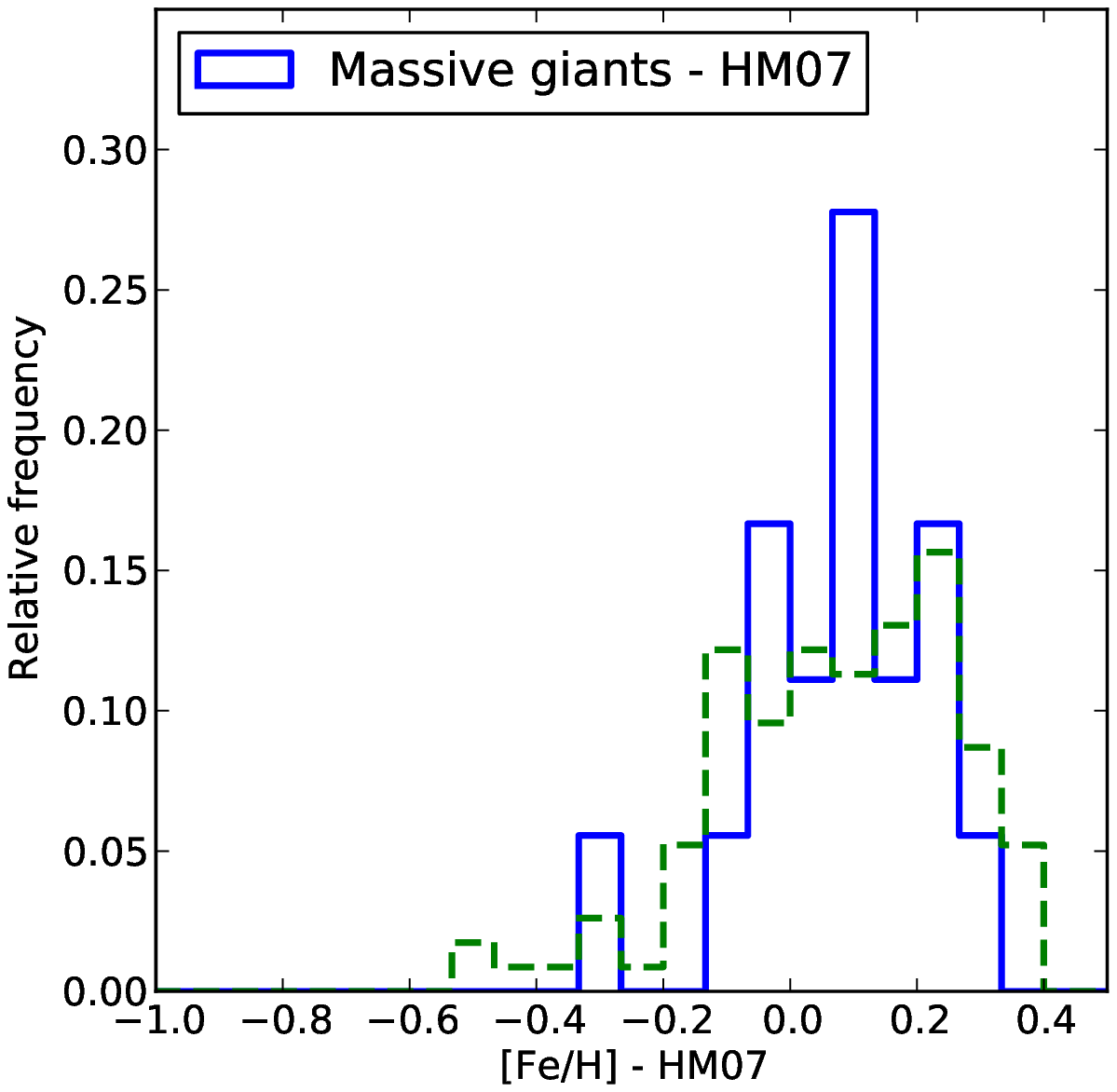}
\caption{Metallicity distribution of evolved stars with giant planets (solid line). The metallicity distribution of the giant-planet host CORALIE + HARPS dwarf sample (dashed line) is shown as a reference. The upper panels use the metallicities derived with the TS13 - SO08 line list set, while the bottom panels use the HM07 line list. From left to right: the complete sample of evolved stars, the subsample of red giants defined through the $\log$g, and the subsample of the most massive giants.}
\label{FigMet}
\end{center}
\end{figure*}

These two metallicity distributions are shown in the top-left panel of Fig. \ref{FigMet}. For the evolved stars, we use our adopted metallicities derived with the TS13 - SO08 line list set. A Kolmogorov-Smirnov test (K-S test) was performed to check whether the two samples follow the same distribution. With a probability of $0.43\%$, we can say with a $3\sigma$ confidence that the evolved sample and the dwarf sample are different in metallicity distribution. All K-S probabilities used in this work are listed in Table \ref{TabKS}.

\begin{table*}
\caption{Probabilities that the two samples are drawn from the same distribution according to a K-S test.}
\label{TabKS}
\centering
\begin{tabular}{ccc}
\hline\hline
Sample 1 & Sample 2 & Probability \\
 & & ($\%$) \\
\hline
Evolved stars ($\log$g $< 4.0$) - TS13 - SO08 & Dwarf planet hosts & $0.43$ \\
Red giants ($\log$g $< 3.0$) - TS13 - SO08 & Dwarf planet hosts & $0.01$ \\
Massive giants (M$_{\ast} > 1.5$ M$_{\odot}$) - TS13 - SO08 & Dwarf planet hosts & $0.28$ \\
Evolved stars ($\log$g $< 4.0$) - HM07 & Dwarf planet hosts & $11.49$ \\
Red giants ($\log$g $< 3.0$) - HM07 & Dwarf planet hosts & $0.01$ \\
Massive giants (M$_{\ast} > 1.5$ M$_{\odot}$) - HM07 & Dwarf planet hosts & $66.49$ \\
Dwarf field stars & Dwarf planet hosts & $8.26\cdot10^{-10}$ \\
Evolved stars ($\log$g $< 4.0$) - TS13 - SO08 & Giant comparison stars & $0.12$ \\
Red giants ($\log$g $< 3.0$) - TS13 - SO08 & Red comparison stars & $27.46$ \\
Evolved stars ($\log$g $< 4.0$) - HM07 & Giant comparison stars & $6.06\cdot10^{-6}$ \\
Red giants ($\log$g $< 3.0$) - HM07 & Red comparison stars & $13.79$ \\

\hline
\end{tabular}
\end{table*}

\begin{figure}[t!]
\begin{center}
\includegraphics[width=6.7cm]{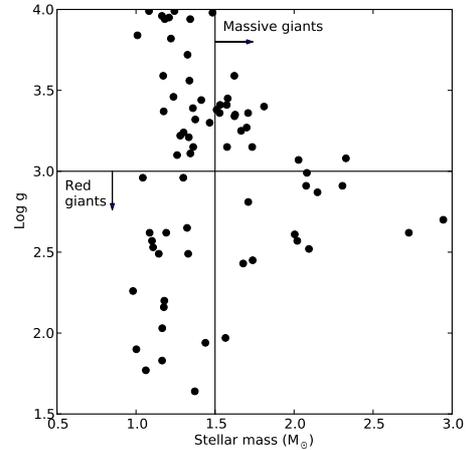}
\caption{Surface gravity $\log g$ versus stellar mass. The values were calculated with the TS13 - SO08 line list set. Solid lines show the boundaries for the different definitions of the giant subsamples.}
\label{FigMass}
\end{center}
\end{figure}

Evolved stars have both lower surface gravities and higher masses than dwarf stars. To test which parameter shows the biggest difference in metallicity distribution for planet hosts, we constructed two subsamples of the evolved sample (where all stars have $\log g < 4.0$). One sample was defined by the surface gravity only ($\log g < 3.0$ - red giants). Another sample was defined by mass only ($M_{\ast} > 1.5 M_{\odot}$ - massive giants). Figure \ref{FigMass} shows the surface gravity versus stellar mass, derived with the TS13 - SO08 line list set. The solid lines represent the boundaries of the two subsamples.

The top-middle and top-right panels of Fig. \ref{FigMet} show the distributions of these two new subsamples for our adopted metallicities. The giant-planet host sample of the CORALIE+HARPS dwarf sample is also shown. We find that the subsample divided by the surface gravity has a different metallicity distribution than the dwarf sample of planet hosts. A K-S test gives a $0.01\%$ probability that the distributions are comparable. When taking the subsample divided by stellar mass only, a K-S test gives a $0.28\%$ probability that the planet host distributions are the same. This result shows that we cannot discard the hypothesis that massive giants with planets and dwarf stars with planets have the same metallicity distribution. Red giants on the other hand do seem to follow a different distribution.

Table \ref{TabMean} lists the mean and median values for all the samples used in this work. Since the samples are not symmetrically distributed in metallicity, we decided to present the trimean\footnote{$T = (Q1 + 2\cdot median + Q3)/4$ where Q1 and Q3 are the first and third quartile.} as well since it is a very good measure of central tendency for unsymmetric distributions. Here we find that the mean metallicity of the red giants, the massive giants, and the planet host dwarfs are respectively -0.179\,dex, -0.024\,dex, and 0.07\,dex. This shows that red giants with planets are on average 0.24\,dex more metal-poor than planet-hosting dwarfs. Using the median or trimean gives similar results.

\begin{table}
\caption{Mean, median, and trimean values of the several samples and subsamples used in this work.}
\label{TabMean}
\centering
\begin{tabular}{c|ccc}
\hline\hline
Sample & Mean & Median & Trimean \\
\hline
Evolved stars - TS13 & -0.055 & 0.0 & -0.020 \\
Red giants ($\log$g $<3.0$) - TS13 & -0.179 & -0.160 & -0.166 \\
Massive giants (mass) - TS13 & -0.024 & 0.020 & 0.004 \\
Evolved stars - HM07 & -0.011 & 0.060 & 0.040 \\
Red giants ($\log$g $<3.0$) - HM07 & -0.180 & -0.210 & -0.198 \\
Massive giants (mass) - HM07 & 0.072 & 0.095 & 0.078 \\
Dwarf stars & -0.106 & -0.080 & -0.085 \\
Planet-hosting dwarf stars & 0.07 & 0.085 & 0.082 \\
Evolved comparison stars  & -0.13 & -0.10 & -0.10 \\
Red comparison giants  & -0.14 & -0.11 & -0.12 \\
\hline
\end{tabular}
\end{table}

Figure \ref{FigMet} also clearly shows that the subsample of red giants defined by the surface gravity has no obvious trend in metallicity. This distribution seems rather flat while the distribution for metallicities from planet hosts in dwarf stars shows a clear increasing trend with increasing metallicity. For the massive giants on the other hand we can see a hint of an increasing trend, similar to the one derived for dwarf stars.

From these tests, it seems that surface gravity is the main parameter responsible for giving different metallicity distributions to planet host stars. Stellar mass also plays a role, but it is less important than the surface gravity.

\subsubsection{Results for the HM07 values}

As discussed in Sect. \ref{CompMet}, the metallicities derived with the TS13 - SO08 line list set may be slightly underestimated for giant stars when compared with other studies. We thus decided to repeat the same analysis presented in the previous section for the metallicities derived with the HM07 line list. The distributions are shown in the bottom panels of Fig. \ref{FigMet}. If we perform a K-S test, we find that there is an $11\%$ probability that the evolved planet hosts and the dwarf planet hosts are drawn from the same distribution. As such, we cannot discard the hypothesis that the samples follow the same distribution. 

For the subsamples (bottom row, middle and right panels of Fig. \ref{FigMet}), we find that the red giants are again differently distributed with only a $0.01\%$ probability that the distribution is the same as the dwarf sample. The massive giants and the dwarf sample seem to follow the same distribution with a K-S probability of $66.49\%$. 

Here we conclude that surface gravity does seem to play a big role in the different metallicity distributions. For stellar mass, the situation is very dependent on which line list is used, but overall it suggests that stellar mass is not the main factor responsible for the observed different metallicity distributions between dwarf planet hosts and evolved planet hosts.

\subsection{Bias in the giant samples}\label{bias}

It is important to note that a comparison between dwarf stars and giant stars suffers from a biased sample selection. Most large programs to search for planets around giant or sub-giant stars select the stars for their sample by making a cut-off in the $B-V$ colour ($B-V\leq1.0$). Examples are the Okayama Planet search program \citep{Sat05}, the retired A stars program \citep{John06}, and the Penn State-Toru\'n Centre for Astronomy Planet Search \citep{Nie08}. The ESO FEROS planet search \citep{Set03a} on the other hand does not perform a $B-V$ cut-off.

\begin{figure}[t!]
\begin{center}
\includegraphics[width=7.0cm]{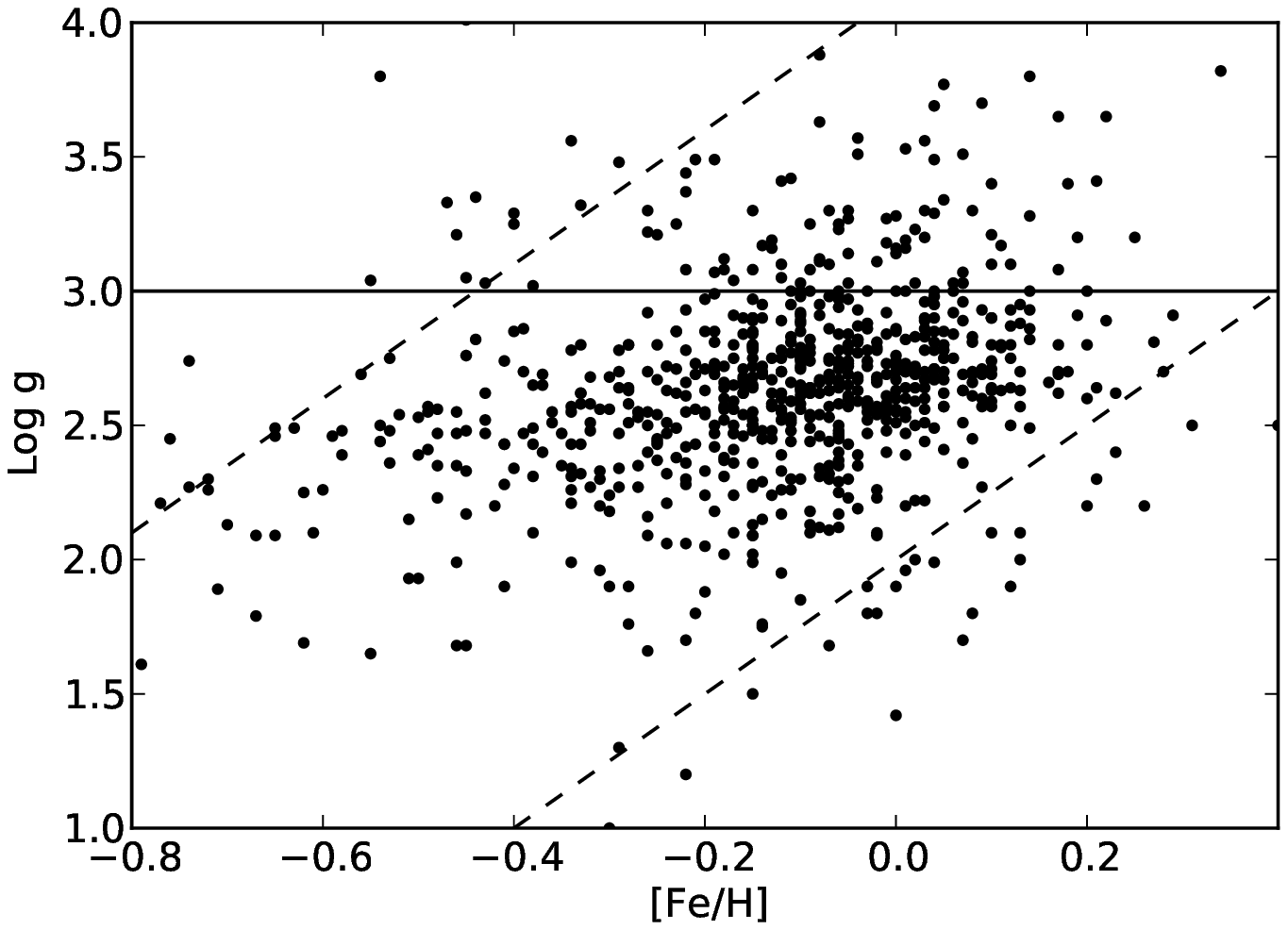}\\
\includegraphics[width=7.0cm]{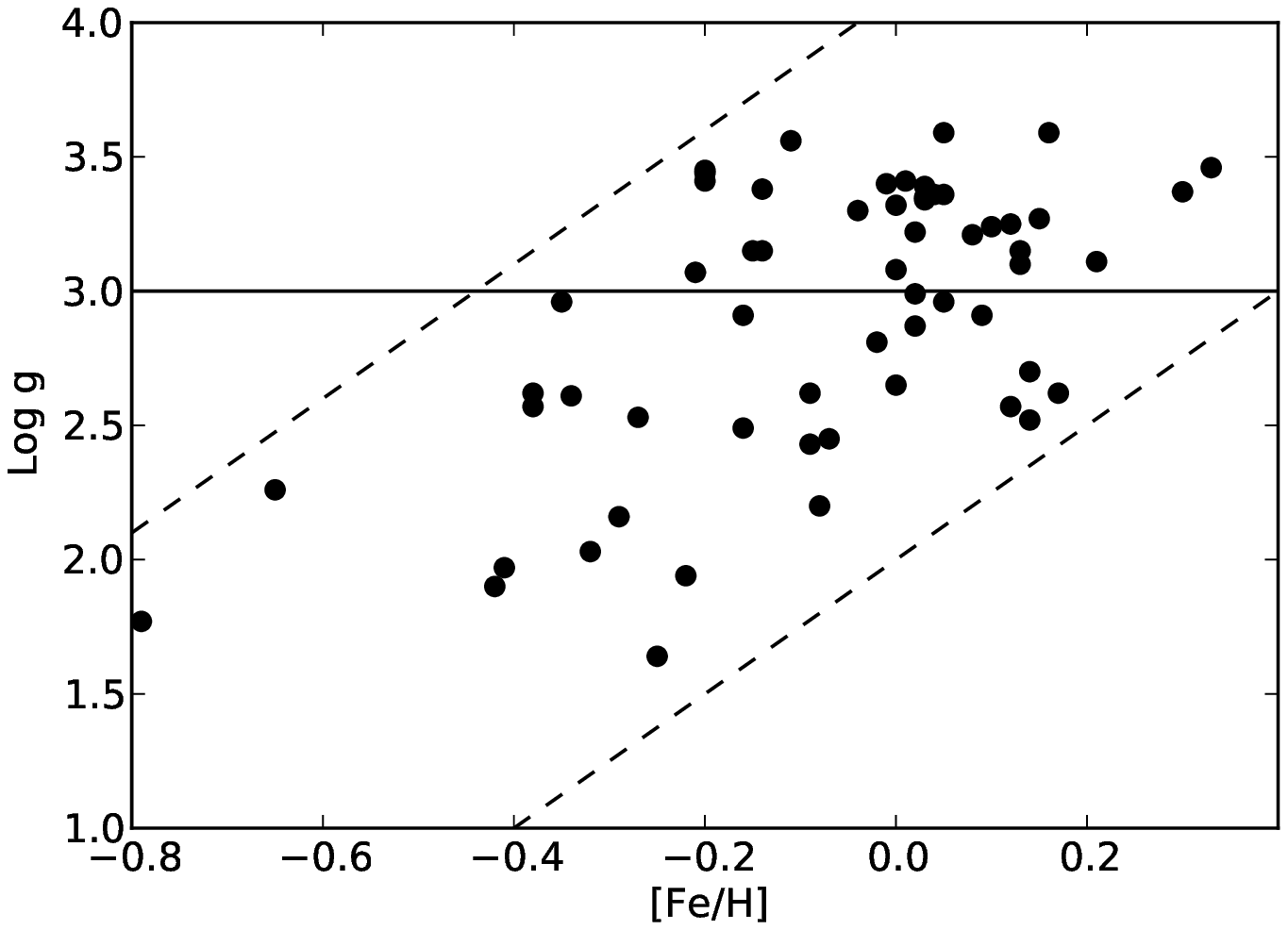}
\caption{Surface gravity $\log g$ versus metallicity for the giant comparison sample used in this work (upper panel) and for our evolved planet hosts (lower panel). The values were calculated with the TS13 line list. The horizontal line denotes the limit in surface gravity for the subsample of red giants. The two dashed lines were drawn by eye and show the biases in the samples due to the B-V cut-off.}
\label{FigLoggMet}
\end{center}
\end{figure}

In Fig. \ref{FigLoggMet} we plot surface gravity $\log g$ versus metallicity [Fe/H]. The top panel shows the stellar parameters for three literature samples. We combined the Okayama Planet search program \citep[parameters from ][]{Tak08}, the ESO FEROS planet search program \citep[parameters from ][]{Das06} and the Penn State-Toru\'n Centre for Astronomy Planet Search \citep[parameters from ][]{Zie12}. For stars that were present in more than one sample, we preferred the parameters from \citet{Tak08} and \citet{Zie12}. All stars have surface gravities lower than 4.0\,dex, compatible with our definition of evolved stars. In the bottom panel we plot our sample of evolved planet hosts. It can be seen from these plots that higher metallicity stars with low surface gravity are left out in these samples. We drew the dashed lines by eye to emphasize this bias.

\begin{figure}[t!]
\begin{center}
\includegraphics[width=6.7cm]{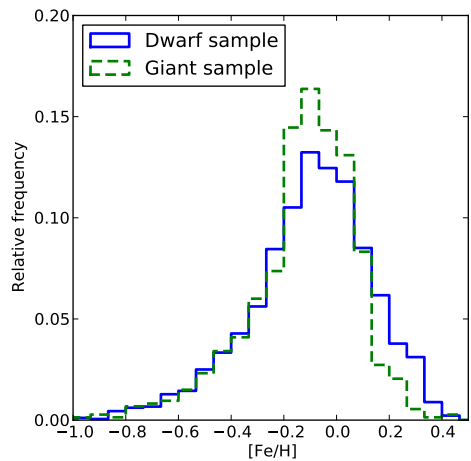}
\caption{Metallicity distribution of the CORALIE+HARPS dwarf sample (solid line) and the giant sample of \citet{Tak08}.}
\label{FigFieldAll}
\end{center}
\end{figure}

For cool stars (which is the case for giant stars), a $B-V$ cut-off results exactly in the lack of high-metallicity, low-gravity stars. The mean temperature of the stars in our evolved sample is about 4850\,K. For this temperature, we calculated the $B-V$ with the calibration of \citet{Sek00}. This calibration depends on temperature, metallicity and surface gravity. Values for different surface gravities and metallicities can be found in Table \ref{TabBV}. Clearly, for low-gravity stars the highest metallicity stars are missed because of the a priori cut-off of $B-V\leq1.0$. In Fig. \ref{FigFieldAll}, we plot the metallicity distributions of both the CORALIE+HARPS dwarf sample (solid line) and the comparison sample of giant stars (dashed line). While the distributions are similar at low metallicities, it is clear that there are fewer giant stars with high metallicity. This is probably a reflection of the $B-V$ cut-off.

\begin{table}
\caption{B-V values for a given temperature of 4850\,K, calculated with the calibration of \citet{Sek00}.}
\label{TabBV}
\centering
\begin{tabular}{c|c|cccc}
\hline\hline
\multicolumn{2}{c|}{} & \multicolumn{4}{c}{[Fe/H]} \\
\cline{3-6}
\multicolumn{2}{c|}{} & -0.2 & 0.0 & 0.2 & 0.4 \\
\hline
 & 1.0 & 0.985 & 1.01 & 1.037 & 1.066 \\
$\log$g & 1.5 & 0.978 & 1.003 & 1.03 & 1.059 \\
 & 2.0 & 0.971 & 0.996 & 1.023 & 1.052 \\
 & 2.5 & 0.964 & 0.989 & 1.016 & 1.044 \\
\hline
\end{tabular}
\end{table}

Since surveys for planets around evolved stars are clearly biased towards lower metallicity, comparisons between evolved stars and dwarf stars with planets should be performed with caution. We found that red giants with planets are on average 0.24\,dex more metal-poor than planet-hosting dwarfs. Given the clear metallicity bias, we recalculated the mean of the planet-hosting dwarfs for stars with metallicities lower than 0.2\,dex. The mean metallicity drops then from $0.07$\,dex to $-0.01$\,dex. Although this mean metallicity is much lower, it is still higher than the mean metallicity for red giants. However, reliable statistics will only be available if we have an unbiased planet search sample of giant stars that includes higher metallicity stars.

\subsection{Planet hosts versus non-planet hosts: a tentative, unbiased comparison}\label{Tak}

To understand whether giant stars with planets also show the metallicity enhancement as observed for dwarf stars with planets, we used the comparison sample of giant stars defined in the previous section. It combines three planet search surveys and consists of 733 giants. Atmospheric parameters were derived with similar methods to this work \citep{Das06,Tak08,Zie12}. We have 19 stars in common with this sample. The atmospheric parameters for these 19 stars are comparable with our results. Mean differences with our atmospheric parameters are -50\,K, -0.11\,dex, and 0.01\,dex for temperature, surface gravity, and metallicity, respectively (see Sect. \ref{Lit}). The mean, median, and trimean metallicity of this sample is shown in Table \ref{TabMean}.

\begin{figure}[t!]
\begin{center}
\includegraphics[width=6.2cm]{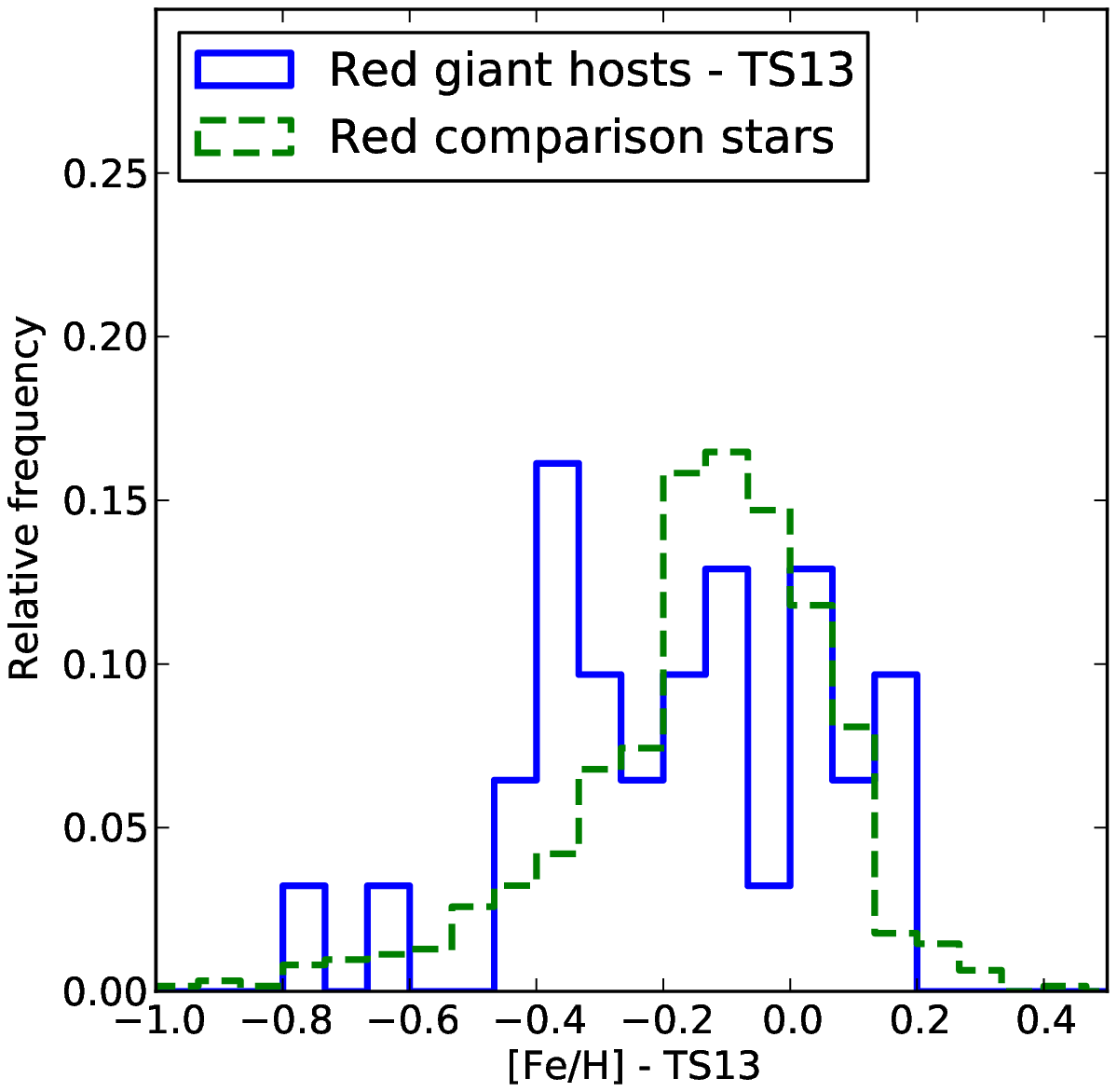}\\
\includegraphics[width=6.2cm]{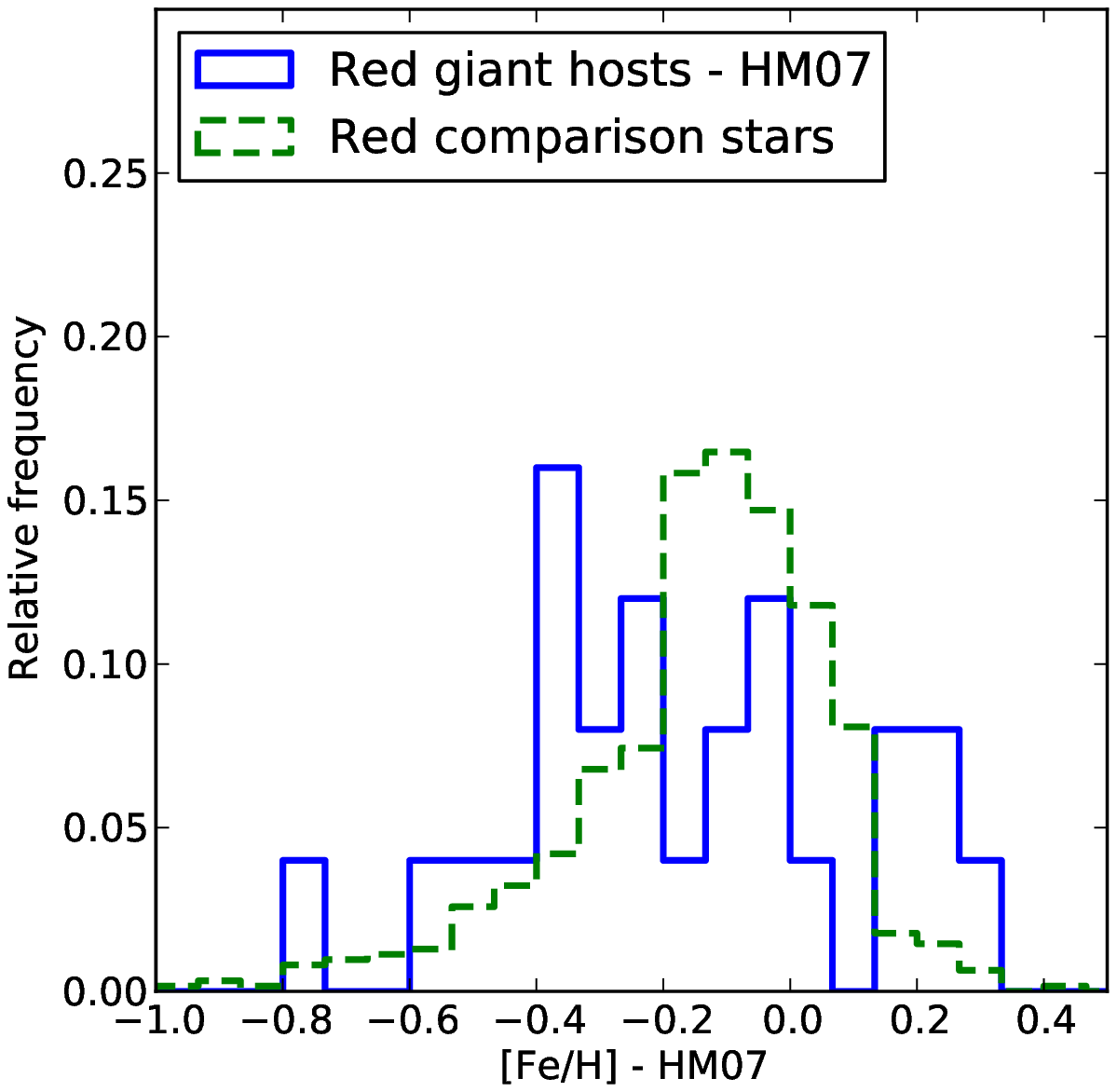}
\caption{Metallicity distribution of red giants. A solid line is used for planet hosts while the dashed line represents the comparison sample of \citet{Tak08}. The upper panel uses the metallicities derived with the TS13 line list. Metallicities in the bottom panel were derived with the HM07 line list. }
\label{FigMetCont}
\end{center}
\end{figure}

For our adopted metallicities derived with the TS13 - SO08 line list set, a K-S test gives a probability of $0.12\%$ that the giant comparison sample and our evolved planet-hosting sample have the same metallicity distribution. In the previous sections, we showed that a different metallicity distribution can be found if we only consider low-gravity stars. In the comparison sample, there are 619 stars with a surface gravity lower than 3\,dex. When comparing this subsample with our subsample of low-gravity planet hosts (see top panel of Fig. \ref{FigMetCont}), the probability that the metallicities follow the same distribution is $27.46\%$. These samples are thus statistically not differently distributed in metallicity. Evolved planet hosts and non-planet hosts follow the same metallicity distribution.

The mean values of the metallicities show that there is a slight metallicity enhancement of about $0.07$\,dex for evolved stars with planets with respect to non-hosting evolved stars (for dwarf stars, the enhancement is $0.17$\,dex). If we consider only the red giants, there is no metallicity enhancement present for planet hosts.

\subsubsection{Results for the HM07 values}

We performed the tests again, also for the parameters derived with the HM07 line list. A K-S test gives a probability of only $6.06\cdot10^{-6}\%$ that the giant comparison sample and this evolved planet-hosting sample have the same metallicity distribution. The lower panel of Fig. \ref{FigMetCont} shows the distributions for the red giant subsamples. The metallicities of red giants with and without planets cannot be statistically distinguished with a K-S probability of $13.79\%$, confirming the result we found with the values derived with the TS13 - SO08 line list set.

The mean values of the metallicities show that there is a metallicity enhancement of about $0.11$\,dex for evolved stars with planets with respect to non-hosting evolved stars. If we consider only the red giants, there is again no metallicity enhancement present for planet hosts. It is clear that the results depend on the line list used, but for the lower gravity samples there is still agreement. No metallicity enhancement can be found for red giant stars with planets.

\section{Discussion}\label{Disc}

We found that evolved ($\log g < 4.0$\,dex) planet hosts are on average 0.24\,dex more metal-poor than planet-hosting dwarfs. This confirms the result of \citet{Pas07}. However, we also found a huge bias in the evolved stellar samples that are being used for detecting planets. High-metallicity red giants are left out because of a cut-off in the $B-V$ colour at 1.0, applied for most of the surveys in search of planets around giant stars.

We find that stellar mass does not play a role in the metallicity distributions of stars with planets. Massive giants and dwarf stars with planets have the same metallicity distribution where the planet frequency increases with metallicity. This is in line with the recent results of \citet{Mal13} who show that massive giants with planets are more metal-rich.

\begin{figure}[t!]
\begin{center}
\includegraphics[width=7.7cm]{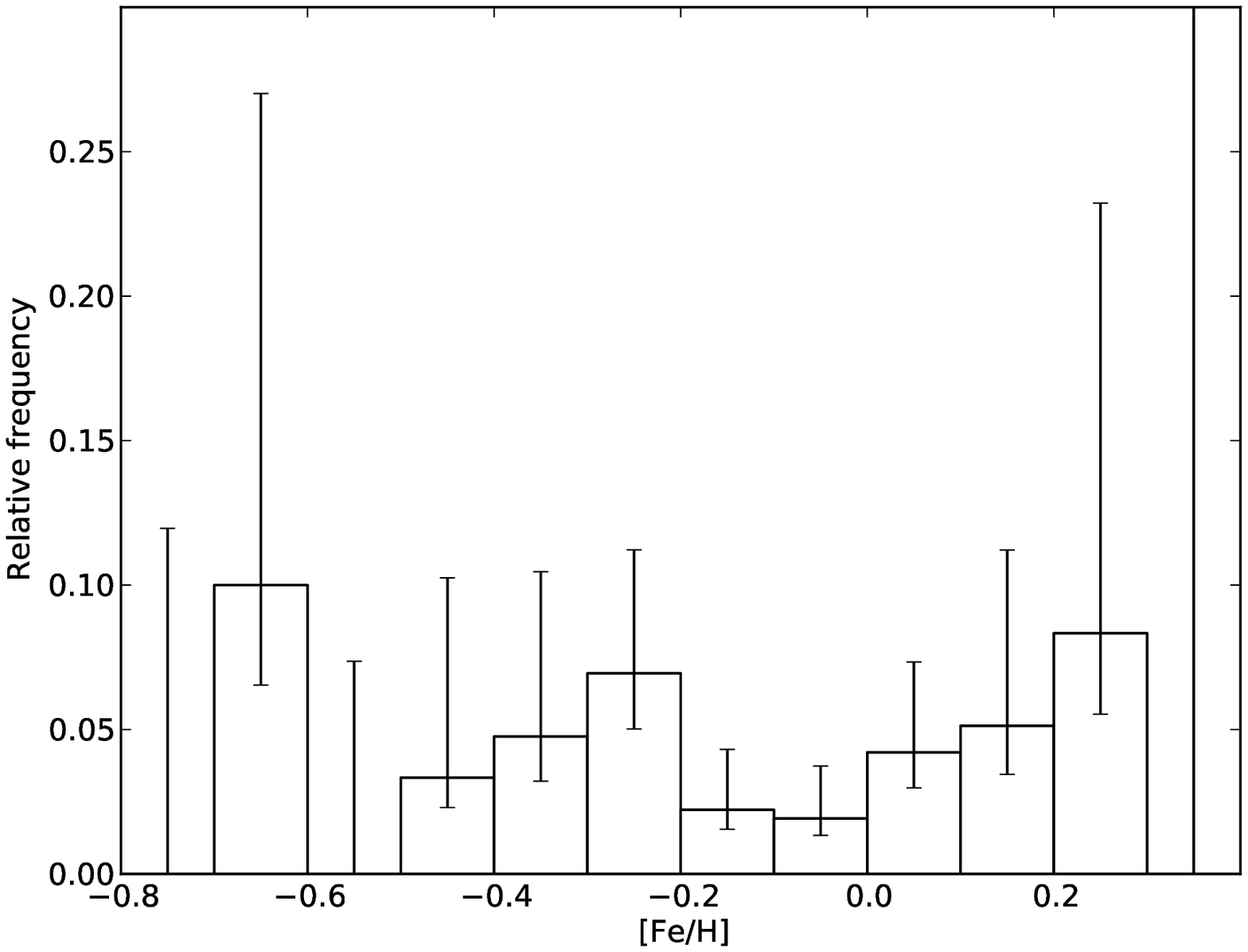}
\caption{Frequency of giant planets as a function of metallicity for the combined red giant sample from \citet{Set03a}, \citet{Sat05}, and \citet{Nie08}.}
\label{FigTak}
\end{center}
\end{figure}

Furthermore, there seems to be no metallicity enhancement present for red giants with planets. Fig. \ref{FigTak} shows the frequency of giant planets as a function of metallicity for the three combined planet-search samples used in this work \citep{Set03a,Sat05,Nie08}. We limited the sample to red giants, with surface gravity $\log$g $<3.0$\,dex. This sample consists of 619 giants, among which 22 planet hosts. The plot shows no clear metallicity enhancement for giant stars with planets. The distribution seems rather flat, although a slight dip can also be seen around solar metallicity. This result reinforces our findings that red giant stars with planets have no metallicity preference. This lack of correlation confirms the results of \citet{Tak08} and \citet{Zie10}. The opposite results from \citet{Hek07} and \citet{Ghe10} could not be confirmed.

The sample of evolved stars with surface gravities between 3.0 and 4.0\,dex (i.e. subgiants) does still show a positive correlation with metallicity. This is probably because these stars are, in terms of evolution, still very close to dwarf stars. They are at an intermediate stage and thus better left out of further discussions.

Since surface gravity and metallicity seem closely linked in red giants, we split the red giant sample into four groups divided at [Fe/H] $= -0.2$ and $\log g = 2.5$, and calculated planet frequencies for these four groups (see Table \ref{TabFreq}). We found that the planet frequency for the low-metallicity, low-gravity stars is $5.55\%$ while the frequency for high-metallicity high-gravity stars is only $2.80\%$. One should not forget, however, that giant stars with the highest metallicity are not being surveyed for planets. If there were no bias in planet survey samples for giant stars, there might be more planet hosts with high metallicity.

\begin{table}
\caption{Giant-planet frequencies with 1-$\sigma$ error bars for four subsamples of red giants. The number of stars in each subsample is shown in brackets.}
\label{TabFreq}
\centering
\begin{tabular}{c|cc}
\hline\hline
 & [Fe/H] $< -0.2$ & [Fe/H] $\geq-0.2$ \\
\hline
$\log g < 2.5$ & $5.55^{+3.09}_{-1.47}\%$  (108) & $3.39^{+2.55}_{-0.98}\%$ (118)\\
$2.5 \leq \log g < 3.0$ & $4.16^{+3.78}_{-1.29}\%$ (72) & $2.80^{+1.25}_{-0.65}\%$ (321)\\
\hline
\end{tabular}
\end{table}

We also caution the reader about these frequency results. We used several survey samples from the literature as a comparison sample. We trust the stellar parameters for the stars in these samples since they compare well with our own results. However, we have no control over the actual survey and the manner in which the search for planets was performed.

Explaining our results is not simple. Several reasons may exist for this lack of metallicity enhancement and for the importance of the surface gravity on the metallicity distribution of evolved planet hosts.

\citet{Pas07} argues that the factor responsible for this lack of correlation is probably the mass of the convective zone. The high metallicity of main-sequence stars is due to the pollution of their atmospheres. In the extended massive envelopes of giant stars, this metal excess is almost completely diluted. This would explain the lack of correlation seen for evolved planet hosts and also the fact that samples of giant stars have fewer metal-rich stars than dwarf star samples. However, subgiants are also diluted and we do see a metallicity correlation for these stars. Furthermore, it seems that this pollution explanation is in contrast with the primordial scenario explanation, where stars are born in high-metallicity clouds \citep[e.g. ][]{Fis05}

Evolved stars are on average more massive than dwarf stars. When stars are more massive, they have more massive proto-planetary disks which may make it easier to form giant planets. As such, the metallicity becomes a less important parameter in the formation process and metal-rich stars would not necessarily be preferred. However, when we select only the most massive stars in our sample, the distribution gets closer to the dwarf sample instead of farther away. The trend towards higher metallicities is also present in this massive subsample. Surface gravity clearly plays a role, but low surface gravities do not necessarily mean higher masses. As such, the stellar mass is probably not the reason for this difference in metallicity.

Another important factor may be the periods of the planets. If metal-rich stars have a greater number of low-period planets (e.g. because migration is faster around these stars), these planets would be engulfed by the star when it becomes a giant (and gets a lower surface gravity). This in turn would result in fewer metal-rich giants with planets. However, when we select only the long-period giants in the CORALIE+HARPS sample, the metallicity enhancement can still be seen for dwarf stars, so it should be present in giant stars as well. This engulfment is thus probably not the reason either.

The metallicity correlation seen in dwarf stars with planets can be explained with the core-accretion formation theory \citep[e.g.][]{Ida04,Udry07,Mor12}. Other planet formation theories, such as gravitational instability, do not expect a trend with metallicity \citep{Boss02}. So, one could argue that planets around evolved stars were formed with another mechanism. However, since all evolved stars were once main-sequence stars, this explanation also seems unfavourable.

To study these giant stars in more depth, there is a need for an unbiased giant sample with no colour cut-off and homogeneously derived parameters that are equally searched for planetary companions. 

\section{Conclusion}\label{Con}

In this paper, we derive atmospheric stellar parameters for a sample of 71 evolved stars with planets. Two different line list sets are considered: a large line list set from TS13 and SO08 for stars cooler and hotter than 5200\,K, respectively, and the small line list from HM07, designed to analyse giant stars. These line lists were tested with the reference star Arcturus which gave good results for all line lists.

We can summarize the results as follows:

\begin{itemize}
\item Surface gravity, microturbulence, and metallicity are not significantly affected by using different line list sets. All values compare well with other spectroscopic results in the literature. The values derived with the TS13 - SO08 line list set are preferred and adopted for further analysis. All these values will be added to SWEET-Cat, a catalogue of stellar parameters for stars with planets \citep{San13}.
\item Using different line list sets provides a very small and constant offset for metallicities that affects the planet frequency statistics. Although the offset is small and within errorbars, it still affects the statistics.
\item Evolved planet hosts are on average 0.24\,dex more metal-poor than planet-hosting dwarfs. There is, however, a strong bias in giant stellar surveys towards lower metallicities. Comparing dwarf stars with giant stars should be done with caution.
\item Only a slight metallicity enhancement is found for evolved stars with planets with respect to evolved stars without planets (depending on the line list used). This enhancement is smaller than the one seen in dwarf stars.
\item No metallicity enhancement is found for red giants with planets with respect to red giants without planets. The metallicity distribution seems flat. Furthermore, the lowest gravity, lowest metallicity stars are slightly preferred for giant planet formation.
\item The reasons for this lack of metallicity enhancement and the preference for lower surface gravities are still unclear. Sample biases cannot be discarded, and a fully uniform study is critical to disentangle the causes of the observed discrepancy.
\end{itemize}

\begin{acknowledgements}
We thank the anonymous referee for the useful comments.
This research has made use of the SIMBAD database, operated at CDS, Strasbourg, France. This work was supported by the European Research Council/European Community under the FP7 through Starting Grant agreement number 239953. E.D.M, S.G.S, and V.Zh.A. acknowledge the support from the Funda\c c\~ao para a Ci\^encia e Tecnologia, FCT (Portugal) in the form of the grants SFRH/BPD/76606/2011, SFRH/BPD/47611/2008, and SFRH/BPD/70574/2010, respectively. G.I. acknowledges financial support from the Spanish Ministry project MICINN AYA2011-29060.
\end{acknowledgements}

\bibliographystyle{aa} 
\bibliography{References.bib}

\appendix

\section{Results from the SO08 line list for cool stars}\label{ApSou}

\begin{figure*}[t!]
\begin{center}
\includegraphics[width=4.5cm]{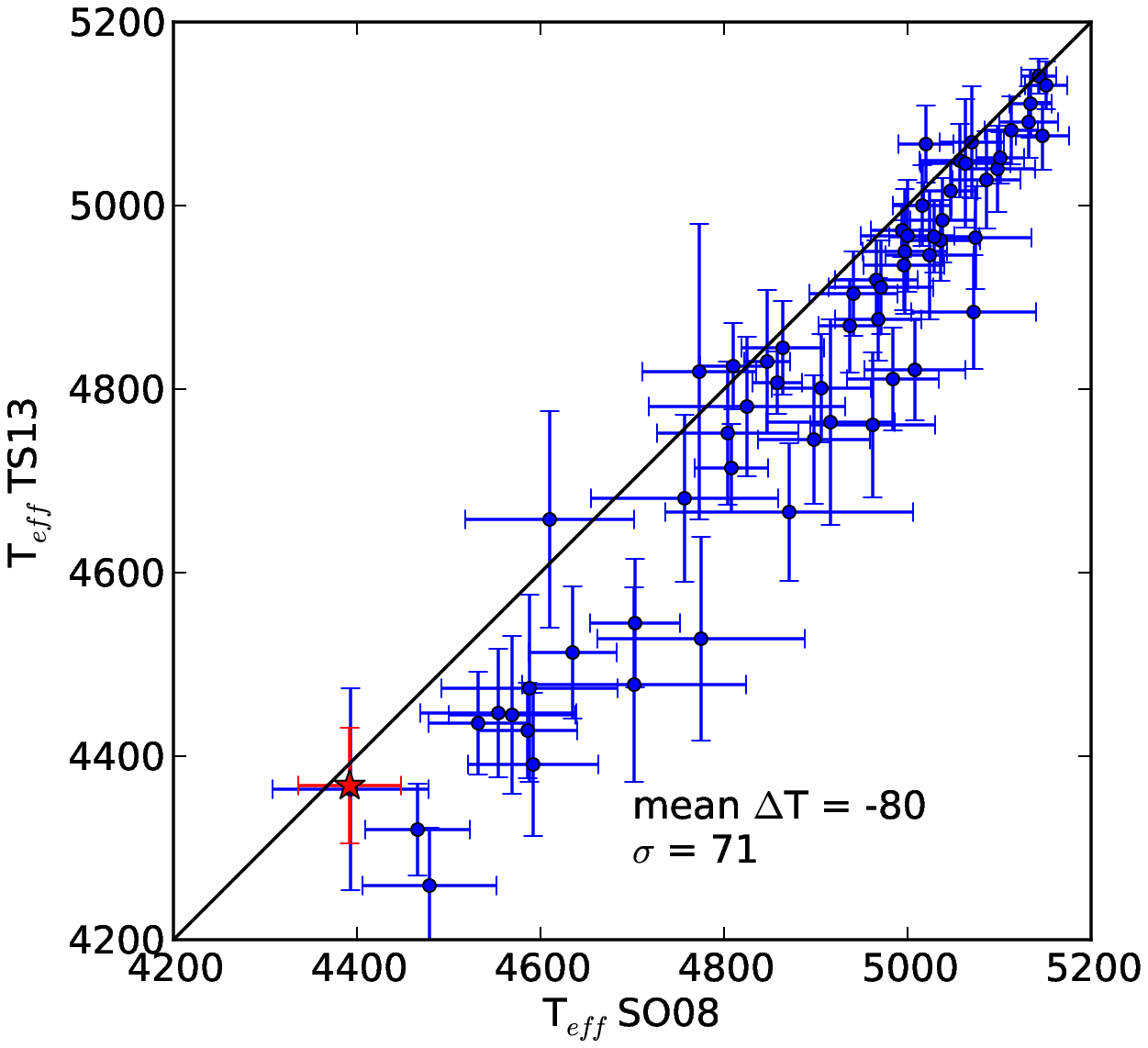}
\includegraphics[width=4.5cm]{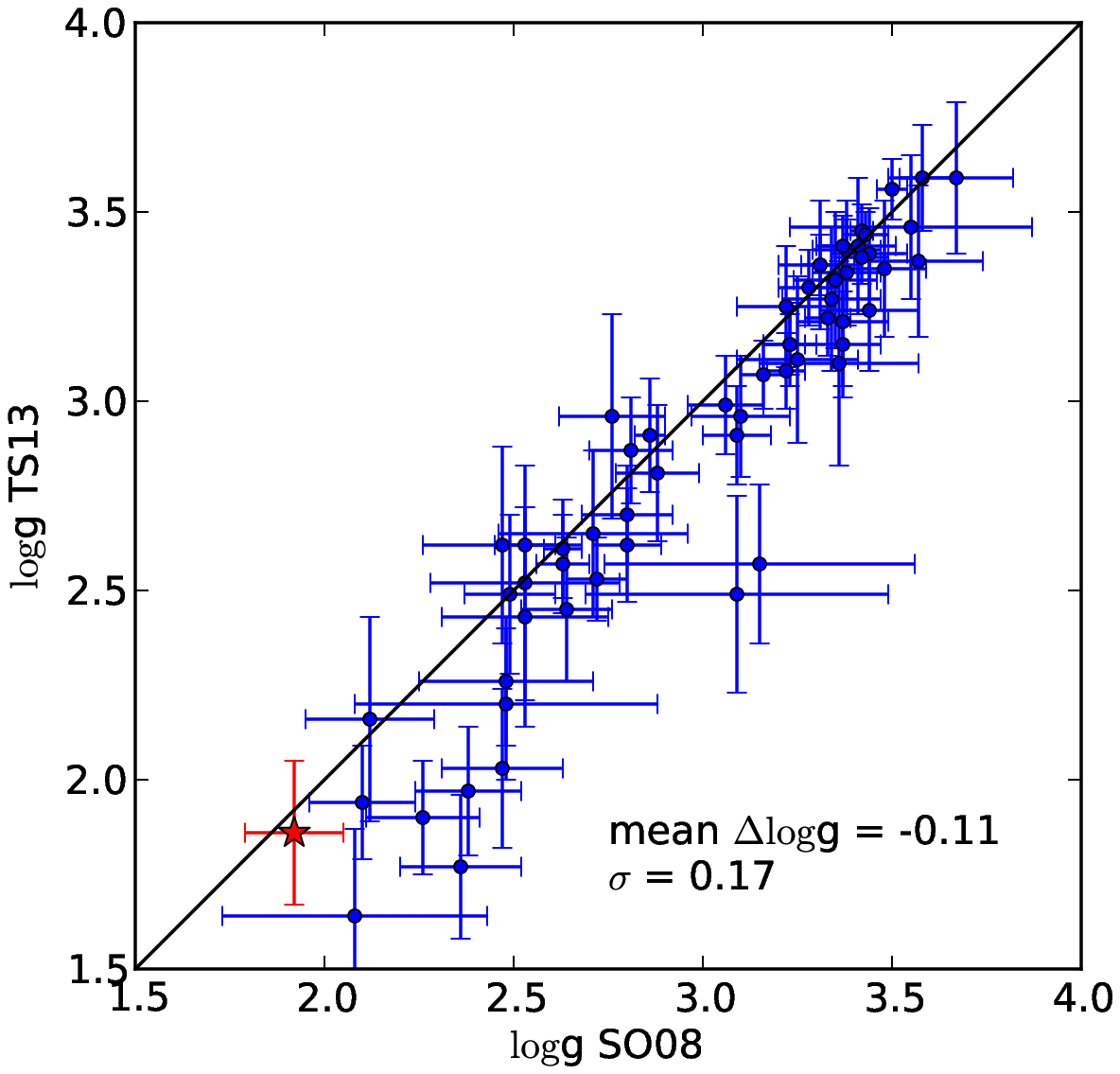}
\includegraphics[width=4.5cm]{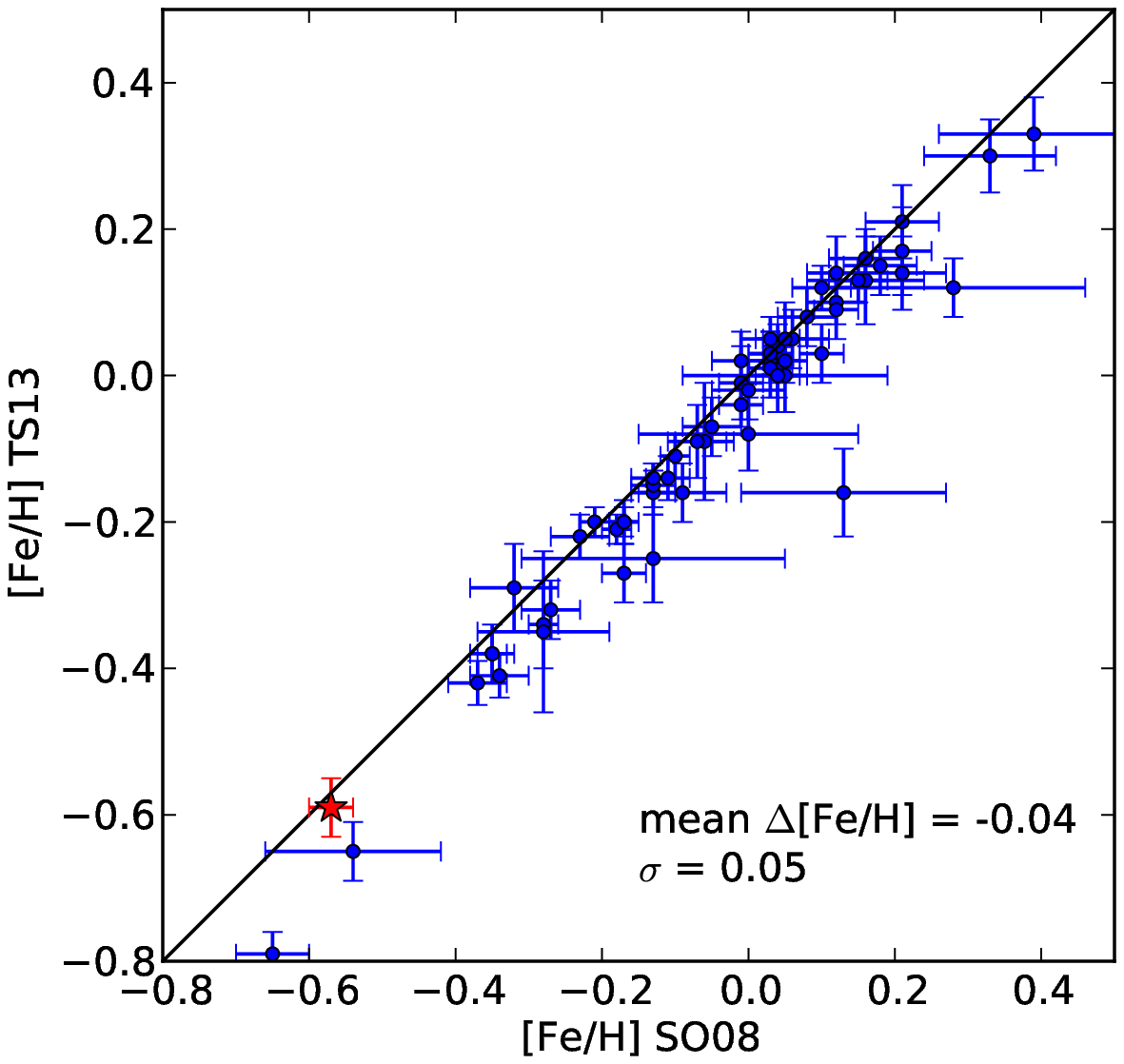}
\includegraphics[width=4.5cm]{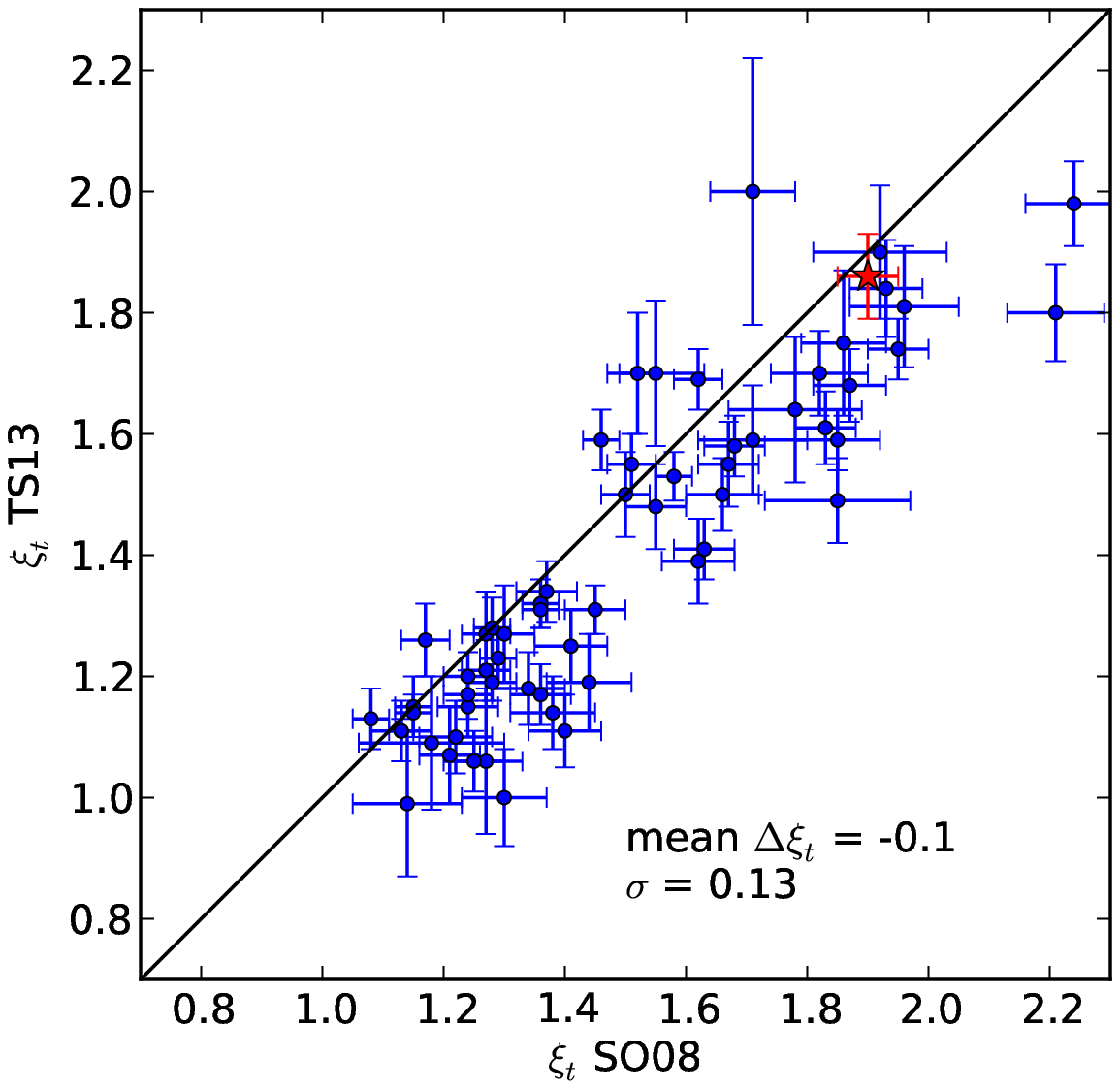}\\
\includegraphics[width=4.5cm]{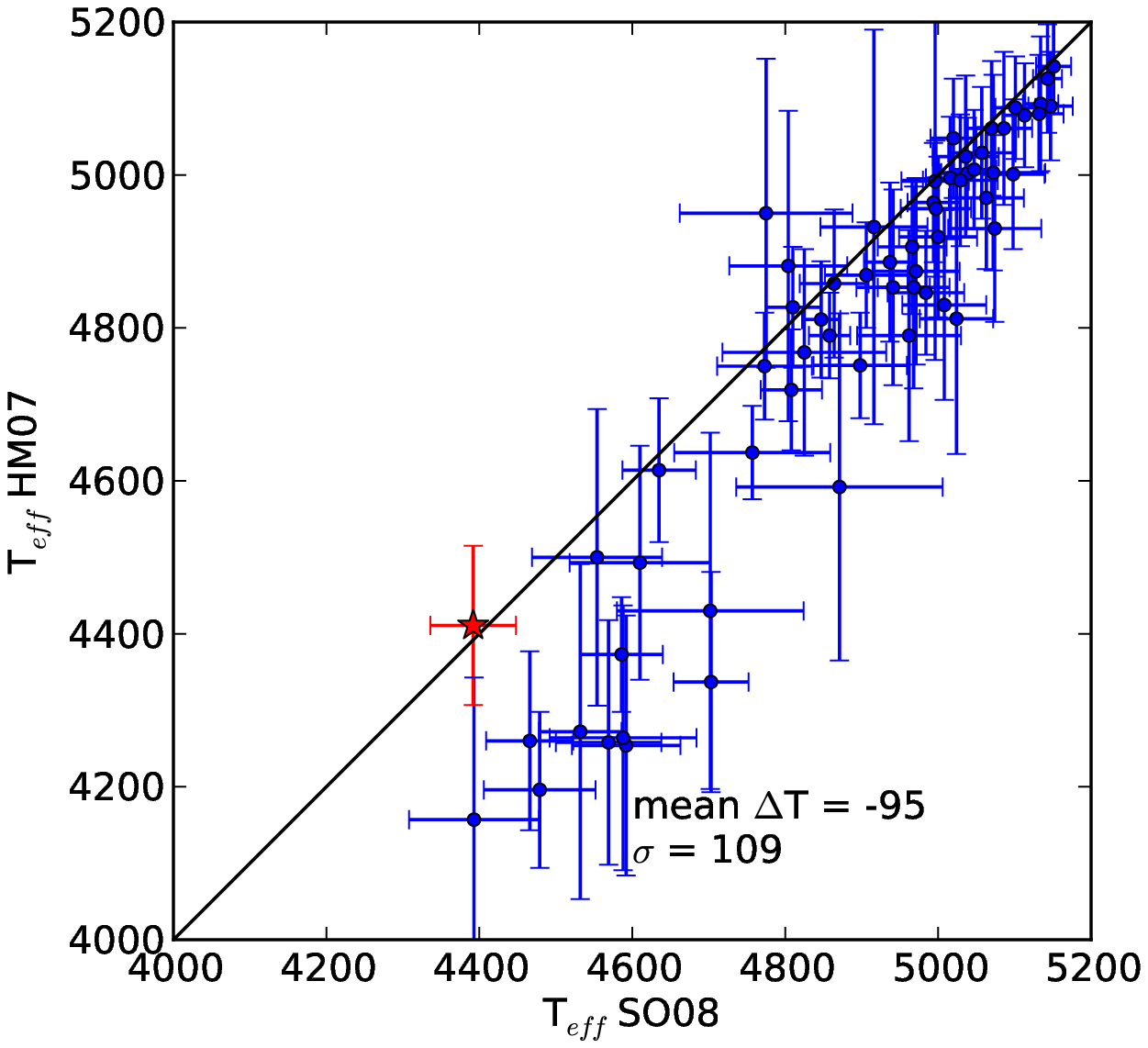}
\includegraphics[width=4.5cm]{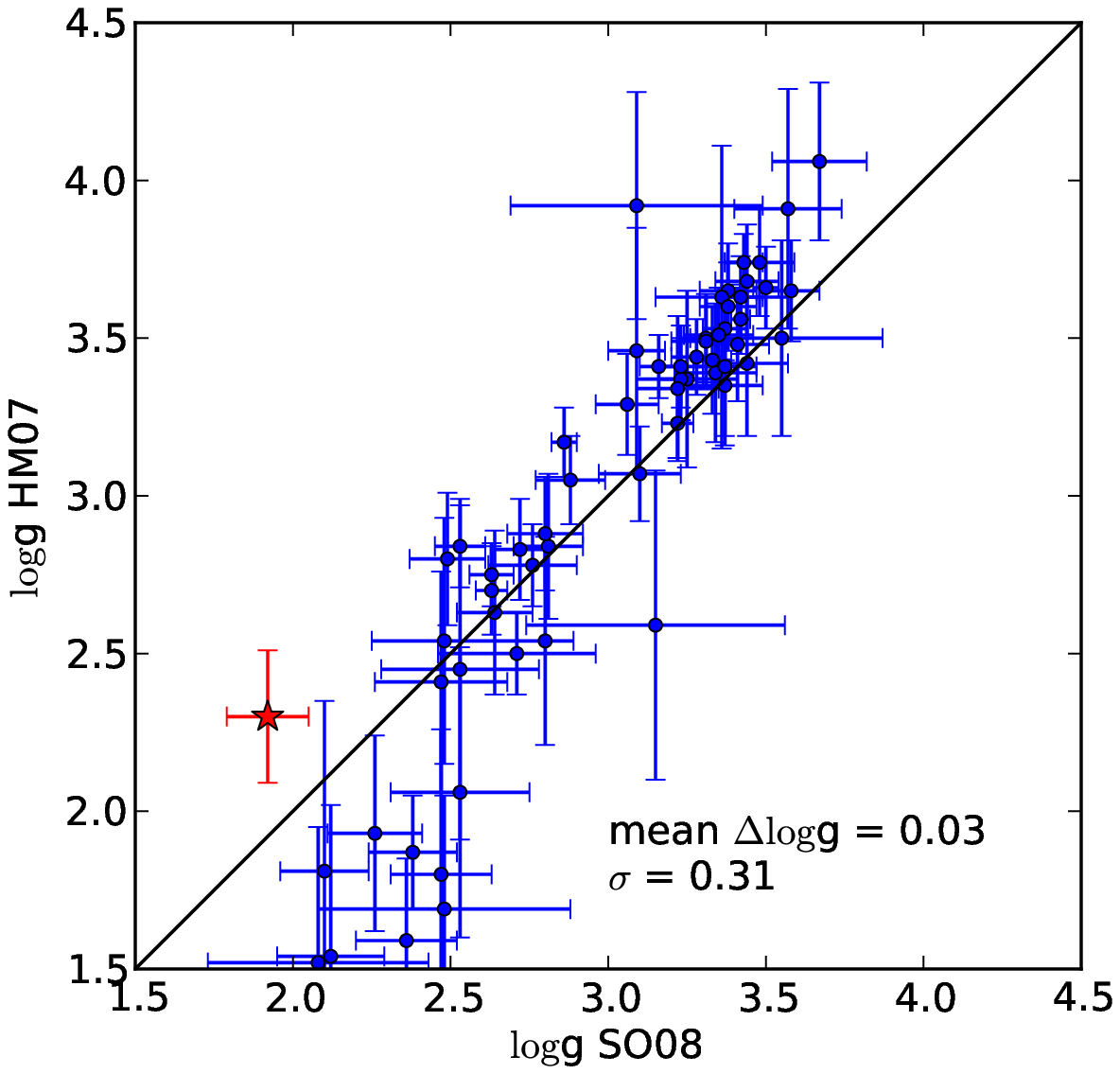}
\includegraphics[width=4.5cm]{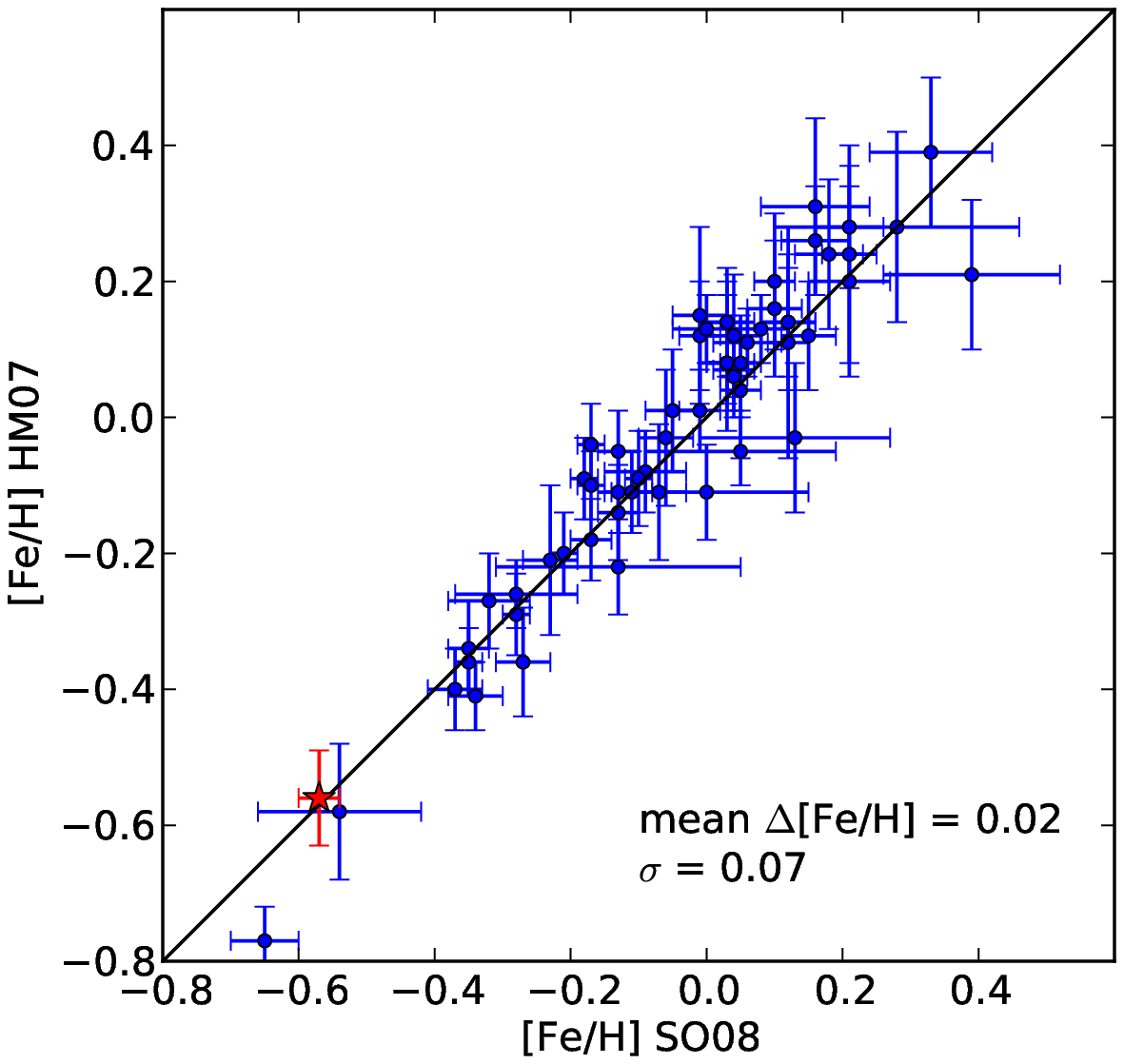}
\includegraphics[width=4.5cm]{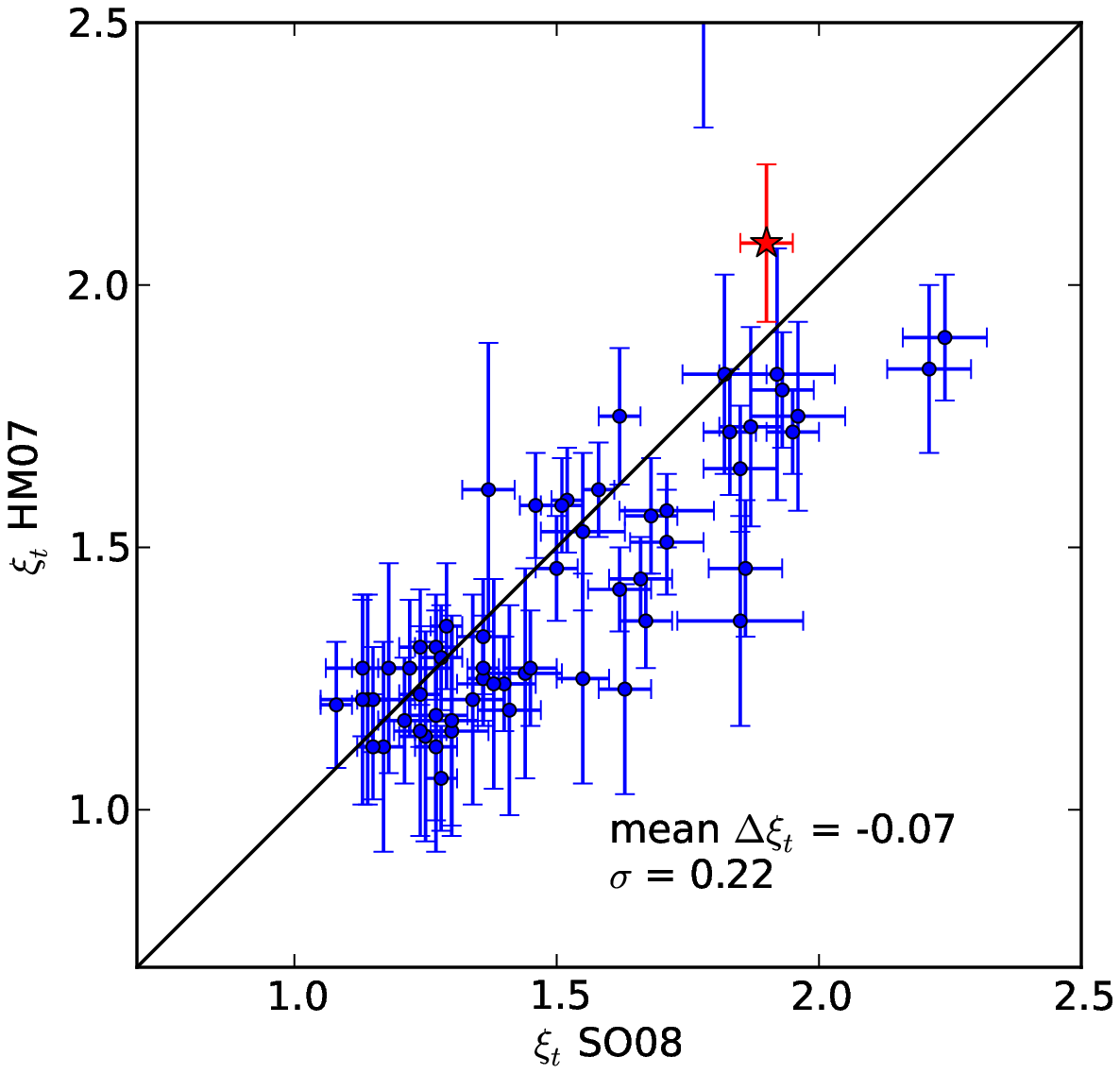}
\caption{Comparisons of the spectroscopic results from the different line lists for effective temperature, surface gravity, metallicity, and microturbulence. In the left panels, the results from the TS13 line list are plotted versus the results from SO08. The middle panels are HM07 versus SO08 and the right panels HM07 versus TS13. The measurements for the reference star Arcturus are overplotted with a star symbol. The dashed line in the top-right panel represents a second degree polynomial fit.}
\label{FigCompSou}
\end{center}
\end{figure*}

Table \ref{TabParSou} lists the results from the SO08 line list for the stars cooler than 5200\,K. Figure \ref{FigCompSou} shows the comparisons of these results with the results from the TS13 and the HM07 line lists.

Effective temperatures for cool stars determined with the SO08 line list have resulted in overestimated values. Because of heavy line blending in the spectra, the equivalent widths of many lines are incorrectly measured, causing incorrect spectroscopic parameters and, specifically, overestimated temperatures \citep{Tsa13}. The TS13 line list has been carefully chosen to resolve this issue. As can be seen in the left panels of Fig. \ref{FigCompSou}, the effective temperatures derived with the SO08 line list are overestimated with respect to the temperatures derived with the TS13 line list and HM07 with a mean difference of $-80$\,K and $-95$\,K, respectively. This difference is higher for the cooler objects, as expected.

The surface gravities determined with the three line lists are comparable with mean differences of $-0.11$ and $0.03$\,dex for TS13 - SO08 and HM07 - SO08, respectively (see second column in Fig. \ref{FigCompSou}). Microturbulences also compare very well with each other (last column in Fig. \ref{FigCompSou}) with mean differences of $-0.10$ and $-0.07$\,km/s, respectively.

Considering metallicities, TS13 found no difference in the results from their line list and the SO08 line list for dwarf stars. Our values confirm these results for evolved stars with a mean difference of $-0.04$\,dex (Col. 3 in Fig. \ref{FigComp}). The metallicities derived with the HM07 line list also compare well with the SO08 metallicities, with a mean difference of $0.02$\,dex.

In general, we confirm the results from \citet{Tsa13}.

\onecolumn
\begin{longtab}
\begin{longtable}{lcccc}
\caption{\label{TabParSou} Stellar parameters derived with the SO08 line list.}\\
\hline\hline
Name & T$_{eff}$ & $\log g_{spec}$ & [Fe/H] & $\xi$ \\
 & (K) & (cm s$^{-2}$) & (dex) & (km s$^{-1}$) \\
\hline
\endfirsthead
\caption{continued.}\\
\hline\hline
Name & T$_{eff}$ & $\log g_{spec}$ & [Fe/H] & $\xi$ \\
 & (K) & (cm s$^{-2}$) & (dex) & (km s$^{-1}$) \\
\hline
\endhead
\hline
\endfoot
11Com & 4847 $\pm$ 25 & 2.63 $\pm$ 0.05 & -0.28 $\pm$ 0.02 & 1.52 $\pm$ 0.03 \\
18Del & 5147 $\pm$ 30 & 3.22 $\pm$ 0.05 & 0.04 $\pm$ 0.03 & 1.36 $\pm$ 0.03 \\
24Sex & 5070 $\pm$ 35 & 3.38 $\pm$ 0.09 & -0.01 $\pm$ 0.03 & 1.27 $\pm$ 0.04 \\
75Cet & 4941 $\pm$ 48 & 2.81 $\pm$ 0.11 & -0.01 $\pm$ 0.04 & 1.63 $\pm$ 0.05 \\
7CMa & 4962 $\pm$ 69 & 3.25 $\pm$ 0.16 & 0.21 $\pm$ 0.05 & 1.44 $\pm$ 0.07 \\
81Cet & 4864 $\pm$ 45 & 2.64 $\pm$ 0.12 & -0.05 $\pm$ 0.04 & 1.68 $\pm$ 0.05 \\
91Aqr & 4757 $\pm$ 102 & 2.71 $\pm$ 0.25 & 0.05 $\pm$ 0.14 & 1.71 $\pm$ 0.09 \\
alfAri & 4635 $\pm$ 49 & 2.49 $\pm$ 0.12 & -0.13 $\pm$ 0.03 & 1.50 $\pm$ 0.04 \\
BD+202457 & 4479 $\pm$ 74 & 2.36 $\pm$ 0.16 & -0.65 $\pm$ 0.05 & 2.24 $\pm$ 0.08 \\
BD+20274 & 4592 $\pm$ 71 & 2.47 $\pm$ 0.16 & -0.27 $\pm$ 0.04 & 2.21 $\pm$ 0.08 \\
BD+48738 & 4610 $\pm$ 92 & 2.47 $\pm$ 0.21 & -0.06 $\pm$ 0.04 & 1.55 $\pm$ 0.08 \\
epsCrB & 4532 $\pm$ 55 & 2.10 $\pm$ 0.14 & -0.23 $\pm$ 0.04 & 1.87 $\pm$ 0.06 \\
epsTau & 5024 $\pm$ 49 & 2.80 $\pm$ 0.09 & 0.21 $\pm$ 0.04 & 1.55 $\pm$ 0.05 \\
gam01Leo & 4586 $\pm$ 55 & 2.38 $\pm$ 0.14 & -0.34 $\pm$ 0.04 & 1.95 $\pm$ 0.05 \\
gammaCephei & 4916 $\pm$ 70 & 3.36 $\pm$ 0.21 & 0.16 $\pm$ 0.08 & 1.27 $\pm$ 0.06 \\
HD100655 & 4906 $\pm$ 55 & 2.88 $\pm$ 0.11 & 0.00 $\pm$ 0.05 & 1.66 $\pm$ 0.06 \\
HD102272 & 4858 $\pm$ 28 & 2.63 $\pm$ 0.07 & -0.35 $\pm$ 0.02 & 1.58 $\pm$ 0.03 \\
HD102329 & 4898 $\pm$ 62 & 3.10 $\pm$ 0.13 & 0.06 $\pm$ 0.05 & 1.62 $\pm$ 0.06 \\
HD104985 & 4773 $\pm$ 62 & 2.76 $\pm$ 0.14 & -0.28 $\pm$ 0.09 & 1.71 $\pm$ 0.07 \\
HD108863 & 4966 $\pm$ 45 & 3.06 $\pm$ 0.10 & 0.03 $\pm$ 0.04 & 1.45 $\pm$ 0.05 \\
HD110014 & 4702 $\pm$ 122 & 2.53 $\pm$ 0.25 & 0.21 $\pm$ 0.06 & 1.92 $\pm$ 0.11 \\
HD116029 & 4984 $\pm$ 50 & 3.37 $\pm$ 0.12 & 0.08 $\pm$ 0.04 & 1.4 $\pm$ 0.06 \\
HD11977 & 5020 $\pm$ 30 & 2.86 $\pm$ 0.04 & -0.09 $\pm$ 0.06 & 1.46 $\pm$ 0.03 \\
HD122430 & 4588 $\pm$ 97 & 2.53 $\pm$ 0.22 & -0.07 $\pm$ 0.04 & 1.96 $\pm$ 0.09 \\
HD13189 & 4337 $\pm$ 133 & 1.83 $\pm$ 0.31 & -0.39 $\pm$ 0.19 & 1.99 $\pm$ 0.12 \\
HD148427 & 5036 $\pm$ 43 & 3.44 $\pm$ 0.10 & 0.03 $\pm$ 0.03 & 1.25 $\pm$ 0.05 \\
HD1502 & 5038 $\pm$ 36 & 3.28 $\pm$ 0.08 & -0.01 $\pm$ 0.03 & 1.27 $\pm$ 0.04 \\
HD167042 & 5086 $\pm$ 38 & 3.48 $\pm$ 0.11 & 0.10 $\pm$ 0.03 & 1.17 $\pm$ 0.04 \\
HD1690 & 4393 $\pm$ 85 & 2.12 $\pm$ 0.17 & -0.32 $\pm$ 0.06 & 1.86 $\pm$ 0.07 \\
HD175541 & 5134 $\pm$ 23 & 3.5 $\pm$ 0.04 & -0.1 $\pm$ 0.02 & 1.08 $\pm$ 0.03 \\
HD177830 & 4804 $\pm$ 77 & 3.57 $\pm$ 0.17 & 0.33 $\pm$ 0.09 & 1.14 $\pm$ 0.09 \\
HD180902 & 5098 $\pm$ 41 & 3.41 $\pm$ 0.10 & 0.03 $\pm$ 0.03 & 1.24 $\pm$ 0.05 \\
HD181342 & 5074 $\pm$ 62 & 3.34 $\pm$ 0.13 & 0.18 $\pm$ 0.05 & 1.41 $\pm$ 0.06 \\
HD18742 & 5047 $\pm$ 30 & 3.23 $\pm$ 0.07 & -0.13 $\pm$ 0.03 & 1.29 $\pm$ 0.03 \\
HD192699 & 5143 $\pm$ 19 & 3.42 $\pm$ 0.03 & -0.21 $\pm$ 0.02 & 1.13 $\pm$ 0.02 \\
HD200964 & 5113 $\pm$ 29 & 3.37 $\pm$ 0.07 & -0.17 $\pm$ 0.02 & 1.28 $\pm$ 0.03 \\
HD206610 & 5008 $\pm$ 56 & 3.44 $\pm$ 0.13 & 0.12 $\pm$ 0.04 & 1.38 $\pm$ 0.07 \\
HD210702 & 5016 $\pm$ 32 & 3.31 $\pm$ 0.05 & 0.04 $\pm$ 0.02 & 1.15 $\pm$ 0.03 \\
HD212771 & 5132 $\pm$ 33 & 3.42 $\pm$ 0.07 & -0.11 $\pm$ 0.03 & 1.24 $\pm$ 0.04 \\
HD27442 & 4825 $\pm$ 107 & 3.55 $\pm$ 0.32 & 0.39 $\pm$ 0.13 & 1.18 $\pm$ 0.12 \\
HD28678 & 5101 $\pm$ 27 & 3.16 $\pm$ 0.06 & -0.18 $\pm$ 0.02 & 1.36 $\pm$ 0.03 \\
HD30856 & 4994 $\pm$ 35 & 3.23 $\pm$ 0.07 & -0.13 $\pm$ 0.03 & 1.28 $\pm$ 0.04 \\
HD33142 & 5057 $\pm$ 45 & 3.38 $\pm$ 0.09 & 0.03 $\pm$ 0.03 & 1.24 $\pm$ 0.04 \\
HD4313 & 5029 $\pm$ 49 & 3.31 $\pm$ 0.11 & 0.03 $\pm$ 0.04 & 1.36 $\pm$ 0.05 \\
HD47536 & 4554 $\pm$ 85 & 2.48 $\pm$ 0.23 & -0.54 $\pm$ 0.12 & 1.82 $\pm$ 0.08 \\
HD5319 & 4937 $\pm$ 34 & 3.33 $\pm$ 0.06 & 0.05 $\pm$ 0.03 & 1.13 $\pm$ 0.05 \\
HD5608 & 4971 $\pm$ 58 & 3.22 $\pm$ 0.13 & 0.1 $\pm$ 0.04 & 1.34 $\pm$ 0.06 \\
HD5891 & 4810 $\pm$ 36 & 2.53 $\pm$ 0.08 & -0.35 $\pm$ 0.03 & 1.62 $\pm$ 0.04 \\
HD59686 & 4871 $\pm$ 135 & 3.15 $\pm$ 0.41 & 0.28 $\pm$ 0.18 & 1.85 $\pm$ 0.12 \\
HD62509 & 4996 $\pm$ 44 & 3.09 $\pm$ 0.09 & 0.12 $\pm$ 0.03 & 1.37 $\pm$ 0.05 \\
HD66141 & 4466 $\pm$ 58 & 2.26 $\pm$ 0.15 & -0.37 $\pm$ 0.04 & 1.85 $\pm$ 0.07 \\
HD73534 & 5072 $\pm$ 69 & 3.67 $\pm$ 0.15 & 0.16 $\pm$ 0.05 & 1.30 $\pm$ 0.07 \\
HD95089 & 4997 $\pm$ 46 & 3.35 $\pm$ 0.11 & 0.04 $\pm$ 0.04 & 1.30 $\pm$ 0.05 \\
HD96063 & 5151 $\pm$ 24 & 3.43 $\pm$ 0.06 & -0.17 $\pm$ 0.02 & 1.15 $\pm$ 0.03 \\
HD98219 & 5063 $\pm$ 49 & 3.58 $\pm$ 0.09 & 0.05 $\pm$ 0.03 & 1.21 $\pm$ 0.05 \\
HIP75458 & 4775 $\pm$ 113 & 3.09 $\pm$ 0.40 & 0.13 $\pm$ 0.14 & 1.78 $\pm$ 0.11 \\
kappaCrB & 4968 $\pm$ 48 & 3.37 $\pm$ 0.10 & 0.15 $\pm$ 0.04 & 1.22 $\pm$ 0.06 \\
ksiAql & 4808 $\pm$ 41 & 2.72 $\pm$ 0.08 & -0.17 $\pm$ 0.03 & 1.51 $\pm$ 0.04 \\
NGC2423 No3 & 4703 $\pm$ 49 & 2.48 $\pm$ 0.40 & 0.00 $\pm$ 0.15 & 1.67 $\pm$ 0.05 \\
NGC4349 No127 & 4569 $\pm$ 69 & 2.08 $\pm$ 0.35 & -0.13 $\pm$ 0.18 & 1.93 $\pm$ 0.06 \\
nuOph & 5000 $\pm$ 51 & 2.80 $\pm$ 0.12 & 0.12 $\pm$ 0.04 & 1.83 $\pm$ 0.05 \\

\end{longtable}
\end{longtab}

\end{document}